\documentclass{article}
\usepackage{latexsym,float}
\usepackage{epsfig,color}
\usepackage{epstopdf}
\usepackage{amssymb}
\usepackage{verbatim}
\usepackage{multirow}
\usepackage{amsmath}
\usepackage{graphicx}
\usepackage{subfig}
\usepackage{epstopdf}
\usepackage{amsmath}
\usepackage{placeins}
\usepackage{subcaption}
\usepackage{graphicx} % Required for inserting images
\usepackage{graphicx}
\usepackage{epstopdf}
\usepackage{amssymb,amsmath,epsfig}
\usepackage{hyperref}
\usepackage{placeins}

\usepackage{xcolor}

\title{\bf Dynamics of Charged Radiating Collapse with Shear and Anisotropy}
\author{
A. Khalid \thanks{ayeshakhalid4238@gmail.com, kha0057@vsb.cz}${}^{~(a,b)}$,
Muhammad Bilal Riaz  \thanks{muhammad.bilal.riaz@vsb.cz, bilalsehole@gmail.com }${}^{~(a)}$,
S. A. Mardan \thanks{syedalimardanazmi@yahoo.com}${}^{~(c,d)}$,\\
Mustafa İnç \thanks{minc@firat.edu.tr}${}^{~(e,f)}$,\\
${}^{a}$IT4Innovations, VSB – Technical University of Ostrava, \\Ostrava, Czech Republic.\\
${}^{b}$ Faculty of Electrical Engineering and Computer Science, \\VSB – Technical University of Ostrava, Ostrava, Czech Republic\\
${}^{c}$ Department of Mathematics, University of Management and Technology,\\
 Lahore, Pakistan.\\
${}^{d}$Center for Theoretical Physics, Khazar University, 41 Mehseti Str.,\\ Baku, AZ1096, Azerbaijan\\
${}^{e}$ Department of Mathematics, Firat University, Elazig, 23119, Turkey\\
${}^{f}$ Department of Mathematics, Khazar University, Baku, Azerbaijan
}

\begin{document}

\maketitle
\begin{abstract}
We investigate a charged anisotropic radiating stellar configuration undergoing gravitational collapse in the presence of shear and heat flux within the Einstein–Maxwell framework. The interior spacetime is described by a time-dependent spherically symmetric geometry and is matched to an exterior charged Vaidya spacetime. The electromagnetic field is incorporated explicitly through Maxwell’s equations, allowing the electric charge to contribute to the matter variables, mass function, and boundary evolution. The charged junction condition is reduced to a Riccati-type differential equation with suitable transformations and exact solution is obtained. The physical properties of the resulting shearing solution are examined through the energy density, radial and tangential pressures, pressure anisotropy, heat flux, electric charge, energy conditions, sound speeds, Herrera cracking criterion, and complexity factor. The energy density and radial pressure remain positive and decrease towards the stellar surface, whereas the tangential pressure remains negative, confirming the anisotropic character of the configuration. Heat transport and electromagnetic effects are strongest in the inner stellar region.  The energy and causality conditions are satisfied. The cracking function indicates potential stability against cracking. The complexity factor remains positive, with electric charge providing an additional contribution alongside pressure anisotropy, density inhomogeneity, and dissipative heat flux. These results provide a comprehensive picture of the physical behavior and internal structure of the charged shearing radiative collapse model.
\end{abstract}

%keywords: Charged; anisotropic fluid; Gravitational collapse; Heat flux; Shear; Complexity factor. 

\section{Introduction}
\label{sec:1}
%--------------Gravitational collapse; 
Gravitational collapse is one of the most fascinating processes in relativistic astrophysics, as it describes how a massive stellar configuration evolves when its internal pressure can no longer balance the inward pull of gravity. The collapse becomes even more physically rich when realistic effects such as electric charge, heat dissipation, pressure anisotropy, and shear are taken into account. Early studies showed that the presence of charge and shear can significantly modify the dynamics of a radiating collapsing star \cite{G1}, while charged shear-free configurations further revealed how electromagnetic effects influence the evolution and stability of the collapsing matter \cite{G2}. A more general treatment of nonadiabatic charged spherical collapse demonstrated that dissipation and electromagnetic fields are strongly coupled to the gravitational evolution of the system \cite{G3}. Later investigations incorporated anisotropy, shear, and heat radiation to construct broader families of charged collapsing solutions \cite{G4}. More recently, attention has shifted to the role of structural complexity in shearing and dissipative collapse, highlighting how different physical ingredients collectively shape the evolution of self-gravitating systems \cite{G5}. These developments make gravitational collapse an active framework for exploring the late stages of stellar evolution under increasingly realistic physical conditions.
\bigskip

%------------------Charged; 
Electric charge and pressure anisotropy are two important ingredients that can noticeably change the structure and evolution of relativistic stellar systems. In a charged configuration, the electromagnetic field introduces an additional repulsive contribution that can counteract gravitational attraction, while anisotropy, characterized by unequal radial and tangential pressures, alters the internal force balance of the matter distribution. The combined influence of these effects has been explored in both dynamical and static stellar models. Rosales et al. \cite{C1} investigated the nonadiabatic evolution of charged spherical systems within the postquasistatic approximation, showing that electromagnetic effects can play a relevant role during stellar evolution. The collapse of charged anisotropic fluids was later studied in detail by Cipolletta and Giambò \cite{C2}, providing a broader description of how charge and anisotropy influence gravitational contraction. Charged anisotropic compact-star models have also been developed using core-envelope configurations, demonstrating that electric charge can contribute to physically viable and stable stellar structures \cite{C3}. The influence of electromagnetic fields on radiating stellar configurations has been examined, further emphasizing the role of charge in dissipative systems \cite{C4}. Solutions of the Maxwell equations for charged anisotropic core-envelope models have likewise strengthened the connection between electromagnetic effects, anisotropic stresses, and the internal properties of compact objects \cite{C5}. These studies highlight that the simultaneous inclusion of charge and anisotropy provides a more flexible and realistic framework for describing highly compact and evolving astrophysical systems.

\bigskip

%--------------Radiating
Alongside charge and pressure anisotropy, radiation and dissipative effects provide another crucial layer in describing the realistic evolution of relativistic stellar systems. During gravitational collapse, a compact object may lose energy through heat flow and radiation, which can significantly alter its internal dynamics and the overall collapse process. In this direction, charged radiating stellar configurations have been investigated by incorporating electromagnetic and dissipative effects simultaneously, leading to physically interesting exact solutions \cite{R1}. The mathematical treatment of radiating matter has also been advanced through new Riccati-type equations, providing useful techniques for constructing wider classes of dissipative stellar models \cite{R2}. Extending these ideas, recent studies have examined radiating stars with nonvanishing shear, demonstrating how shear stresses can further modify the dynamics of collapsing matter \cite{R3}. Anisotropic shear-free radiating collapse has likewise been explored through exact solutions and numerical validation, highlighting the close interplay among anisotropy, heat transport, and stellar evolution \cite{R4}. Radiating configurations have also been extended beyond general relativity, including models formulated within Einstein--Gauss--Bonnet gravity \cite{R5}. Taken together, these developments show that charge, anisotropy, radiation, and shear should not be viewed as isolated effects; instead, their combined presence provides a more complete framework for studying the complex physics of gravitational collapse.

\bigskip

%--------------Heat flux; 
It is not surprising that one of the important ways of transporting energy from the interior to the exterior of a collapsing star is through heat flux in this overall dissipative setting. It can have a direct impact on the temperature profile, pressure distribution, stability and overall rate of stellar evolution. The thermal behaviour of radiating anisotropic stars with shear has been studied in detail and it has been shown that the heat transport can play an important role in the internal dynamics of a collapsing configuration \cite{H1}. The effects of heat dissipation, electric charge, shear and viscosity have also been investigated and it has been demonstrated that all these effects can interact to modify the collapse process and the evolution of the stellar matter \cite{H2}. The equation of state has been found to be an important factor in determining the dynamical stability of radiating stars, especially in the presence of heat flow \cite{H3}. The dissipative effects have also been linked to the formation of structural complexity in relativistic collapse, and heat transport is closely related to the internal organization of self-gravitating systems \cite{H4}. Thus, in the realm of charge, anisotropy and radiation, it is natural and physically motivated to add heat flux to the development of a more complete description of realistic gravitational collapse.
\bigskip

%--------------Shear; 
Closely connected with heat transport, shear is another important dynamical feature that can significantly reshape the evolution of a radiating and collapsing stellar configuration. It describes the unequal deformation of neighboring fluid layers and therefore becomes especially relevant when different regions of the star evolve at different rates. Shearing radiative collapse with expansion and acceleration has been studied to clarify how these kinematical effects influence the temporal development of relativistic matter distributions \cite{S1}. Later investigations further examined the time-dependent behavior of radiating stars, providing deeper insight into how their internal geometry and matter variables evolve during collapse \cite{S2}. The combined presence of shear and heat dissipation has also been linked with the complexity of self-gravitating systems, showing that these effects can strongly influence the internal organization of the collapsing fluid \cite{S3}. The studies on shearing radiative collapse, reveals that the shear can introduce richer dynamical behavior and multiple evolutionary pathways in relativistic stellar models \cite{S4}. Hence, following the roles of charge, anisotropy, radiation, and heat flux, the inclusion of shear completes a more realistic dynamical picture and provides a natural foundation for studying the collective impact of these physical effects on gravitational collapse.

\bigskip

%--------------Complexity factor.
With charge, anisotropy, radiation, heat flux, and shear all shaping the evolution of a collapsing star, a natural question is how these effects can be brought together within a single measure of the system’s internal structure. This motivation leads to the concept of the complexity factor, which provides a useful way to characterize the combined influence of density inhomogeneity, pressure anisotropy, dissipation, and gravitational effects in self-gravitating systems. Herrera et al. \cite{Y1} introduced a definition of complexity for dynamical spherically symmetric dissipative fluids, establishing a systematic framework for studying how different physical sources contribute to the structural complexity of evolving stellar configurations. The role of electric charge was later examined in connection with cracking and complexity, showing that electromagnetic effects can significantly modify the internal balance of dissipative compact objects \cite{Y2}. The impact of shear on the complexity factor and Weyl stresses has also been explored during dissipative collapse, further highlighting the close relation between kinematical deformation and the gravitational structure of the fluid \cite{Y3}. Charged compact stellar models under the zero-complexity condition have additionally been investigated, demonstrating how this constraint can be used to construct and analyze physically viable configurations \cite{Y4}. More recently, the combined influence of electric charge, complexity, and Weyl stresses has been studied in radiating gravitational collapse, bringing these ideas closer to a unified description of dynamical charged systems \cite{Y5}. Therefore, the complexity factor provides a natural bridge connecting the different physical ingredients considered above and offers a compact way to investigate how they collectively govern the evolution of relativistic stellar collapse.

\bigskip

Electric charge is introduced to analyze the influence of electromagnetic forces on the dynamics of a radiative shear flow collapse, but not considering charge capable of stabilizing the entire system. This work is structured as follows: Sec. \ref{sec:2} describes the charged star model and its field equations, Sec. \ref{sec:3} obtains the exact solution, Sec. \ref{sec:4} analyzes physical and stability issues, Sec. \ref{sec:5} focuses on anisotropy and complexity, while Sec. \ref{sec:6} draws the conclusions.

\bigskip

\section{The model}
\label{sec:2}
In order to describe the collapsing configuration consistently in terms of charge, anisotropy and radiation, we first define the geometrical and matter sectors of the system. The formulation is broken down into three interrelated sections. The interior spacetime and the characteristics of the fluid are introduced first. Then, the electromagnetic contribution is added in using the electromagnetic energy–momentum tensor and Maxwell's equations. Lastly, the gravitational and matter variables are connected by the Einstein field equations. This is a system in which the effects of anisotropy, heat flux, shear, and electric charge can be dealt with in a unified dynamical framework.
\bigskip

\subsection{Interior spacetime}
We consider a time-dependent spherically symmetric interior spacetime described by the line element

\begin{equation}\label{eq:1}
ds_{-}^{2}=-A^{2}(r,t)dt^{2}+B^{2}(r,t)dr^{2}
+Y^{2}(r,t)\left(d\theta^2+\sin^2\theta\, d\phi^2\right),
\end{equation}

with $A(r,t)$ and $B(r,t)$ being the temporal and radial metric potentials, respectively, while $Y(r,t)$ represents the areal radius of the collapsing system. Given that these quantities are dependent upon both $r$ and $t$, the geometry automatically accounts for the dynamic behavior of the star's interior. The geometrical quantities are expressed through the comoving fluid four-velocity $u^{a} = \frac{1}{A}\delta^{a}_{0}$, acceleration vector $\dot u^{a}$, the expansion
scalar is $\Theta$, and and the shear scalar is $\sigma$. In spherical symmetry they are

\begin{subequations}\label{eq:2}
\begin{align}
u^{a} &= \left(1/A,\,0,\,0,\,0\right), \tag{2a}\label{eq:2a}
\\[4pt]
\dot{u}^{a} &= \left(0,\,\frac{A'}{A B^{2}},\,0,\,0\right), \tag{2b}\label{eq:2b}
\\[4pt]
\Theta &= \frac{1}{A}
\left(\frac{\dot{2Y}}{Y}+
\frac{\dot{B}}{B}
\right), \tag{2c}\label{eq:2c}
\\[4pt]
\sigma &= 
\left(\frac{\dot{Y}}{Y}-
\frac{\dot{B}}{B}
\right)\frac{1}{3A}, \tag{2d}\label{eq:2d}
\end{align}
\end{subequations}

here, an overdot and a prime denote differentiation to $t$ and $r$. The energy momentum tensor for the inner matter is

\begin{equation}\label{eq:3}
T^{-}_{ab}=(\rho +p_{t})u_{a}u_{b}+p_{t}g_{ab}
+(p_{r}-p_{t})n_{a}n_{b}+q_{a}u_{b}+q_{b}u_{a}.
\end{equation}

Where $\rho$ is the density, $p_{r}$ and $p_{t}$ are the radial and tangential pressures, four-velocity is $u^{a}$, $n^{a}$ is the radial unit spacelike vector, and $q^{a}$ represents the heat-flux vector. The difference $p_{t}-p_{r}$ accounts for local pressure anisotropy, while the heat-flux terms describe the outward transport of energy from the stellar interior. The vectors $u^{a}$, $n^{a}$, and $q^{a}$ satisfy

\begin{equation}\label{eq:4}
u^{a}u_{a}=-1, \qquad n^{a}n_{a}=1, \qquad u^{a}n_{a}=0, \qquad q^{a}u_{a}=0.
\end{equation}

This provide correct normalization of both timelike and spacelike directions and that make the heat flux spatial in the comoving coordinates frame. This set of geometrical and physical quantities serves as the base for the further introduction of electromagnetic field sector and construction of full set of Einstein–Maxwell equations.

\subsection{The electromagnetic energy tensor and the Maxwell
equations}
The electromagnetic energy tensor $E^{-}_{ab}$ is given by 

\begin{equation}\label{eq:5}
E^{-}_{ab}=\frac{1}{4\pi}\left(F_{a}^{c}F_{bc}-\frac{1}{4}F^{cd}F_{cd}g_{ab}\right),
\end{equation}

where $F_{ab}$ is the electromagnetic field tensor and $g_{ab}$ denotes the spacetime metric. The field tensor is generated from the electromagnetic four-potential $\phi_a$ according to

\begin{equation}\label{eq:6}
F_{ab}=-\phi_{a,b}+\phi_{b,a}.
\end{equation}

The electromagnetic field and the motion of the charged matter are connected through Maxwell's equation

\begin{equation}\label{eq:7}
F^{ab}{}_{;b}=4\pi J^{a},
\end{equation}

where $J^{a}$ is the four-current density. For a spherically symmetric charged distribution, we take

\begin{equation}\label{eq:8}
\phi_{a}=\Phi(r,t),\delta^{0}_{a},\qquad J^{a}=\varsigma V^{a},
\end{equation}

where $\Phi(r,t)$ is the electric potential and $\varsigma$ is the proper charge density. Charge conservation implies that

\begin{equation}\label{eq:9}
\mathcal{Q}(r)=4\pi\int_{0}^{r}
\varsigma\,B\,Y^{2}\,dr.
\end{equation}

Using the metric (\ref{eq:1}) and equations (\ref{eq:6}-\ref{eq:7}), we obtain
\begin{equation}\label{eq:10}
\Phi''-\left(\frac{B'}{B}+\frac{A'}{A}-2\frac{Y'}{Y}\right)\Phi'=4\pi\varsigma A B^{2}.
\end{equation}

\begin{equation}\label{eq:11}
\dot{\Phi}'-\left(\frac{\dot{B}}{B}+\frac{\dot{A}}{A}-2\frac{\dot{Y}}{Y}\right)\Phi'=0.
\end{equation}

Integrating the above equations gives
\begin{equation}\label{eq:12}
\Phi'=\frac{\mathcal{Q}(r)AB}{Y^{2}},
\end{equation}

which directly links the electric field with the charge enclosed inside the star and the geometry of the interior spacetime.

\subsection{The field equations}

The matter and electromagnetic parts are specified; we now combine them for the description of the full dynamics of the gravitational collapse in the charged case. The field equations are expressed as

\begin{equation}\label{eq:13}
G^{-}_{ab}=8\pi\left(T^{-}_{ab}+E^{-}_{ab}\right),
\end{equation}

where $G^{-}_{ab}$ denotes the Einstein tensor, and $T^{-}_{ab}$ and $E^{-}_{ab}$ are the energy–momentum tensors for matter and electromagnetism, respectively. Hence, the spacetime geometry of the interior solution depends on both the fluid elements and the electric field. From equations (\ref{eq:3}), (\ref{eq:5}) and (\ref{eq:13}), the system of field equation becomes

\begin{subequations}\label{eq:14}
\begin{align}
8\pi \rho + \frac{\mathcal{Q}^{2}}{Y^{4}}&=
\frac{2}{A^{2}}\frac{\dot{B}\dot{Y}}{BY}
+\frac{1}{A^{2}}\frac{\dot{Y}^{2}}{Y^{2}}
+\frac{1}{Y^{2}}
-\frac{1}{B^{2}}
\left(
\frac{Y'^2}{Y^{2}}
+2\frac{Y''}{Y}
-2\frac{B'Y'}{BY}
\right),
\tag{14a}\label{eq:14a}
\\[6pt]
8\pi p_{r} -\frac{\mathcal{Q}^{2}}{Y^{4}}&=
\frac{1}{A^{2}}
\left(
-2\frac{\ddot{Y}}{Y}
+2\frac{\dot{A}\dot{Y}}{AY}
-\frac{\dot{Y}^{2}}{Y^{2}}
\right)
+\frac{1}{B^{2}}
\left(
2\frac{A'Y'}{AY}
+\frac{Y'^2}{Y^{2}}
\right)
-\frac{1}{Y^{2}},
\tag{14b}\label{eq:14b}
\\[6pt]
8\pi p_{t} +\frac{\mathcal{Q}^{2}}{Y^{4}}&=
-\frac{1}{A^{2}}
\left(
\frac{\ddot{B}}{B}
-\frac{\dot{A}\dot{B}}{AB}
+\frac{\dot{B}\dot{Y}}{BY}
+\frac{\ddot{Y}}{Y}
-\frac{\dot{A}\dot{Y}}{AY}
\right)
\nonumber\\
&\quad
+\frac{1}{B^{2}}
\left(
\frac{A''}{A}
+\frac{A'Y'}{AY}
-\frac{A'B'}{AB}
+\frac{Y''}{Y}
-\frac{B'Y'}{BY}
\right),
\tag{14c}\label{eq:14c}
\\[6pt]
8\pi Q_{h} &=
-\frac{2}{AB^{2}}
\left(
-\frac{\dot{Y}'}{Y}
+\frac{A'}{A}\frac{\dot{Y}}{Y}
+\frac{\dot{B}}{B}\frac{Y'}{Y}
\right).
\tag{14d}\label{eq:14d}
\end{align}
\end{subequations}

The term $\mathcal{Q}^{2}/Y^{4}$ represents the electromagnetic contribution, while the remaining terms are generated by the radial and temporal behavior of the metric functions. The discrepancy between $p_r$ and $p_t$ is due to the anisotropy property of the stellar material, whereas the charge terms reveal the contribution of the electromagnetic field in different directions. In this context, $Q_{h}=qB$ and $q^{a}=(0,q,0,0)$ stands for the heat flow function along the radial direction. \\

The inner spacetime (\ref{eq:1}) has to match smoothly at star's surface to the exterior spacetime (Vaidya) that is radiating, and given by the metric

\begin{equation}\label{eq:15}
ds_{+}^{2}=-\left(1-\frac{2m(v)}{\tilde r}+\frac{\mathcal{Q}^{2}}{\tilde r^{2}}\right)dv^{2}-(2\,dv\,d\tilde r) +\tilde r^{2}(d\theta^2+\sin^2\theta\, d\phi^2),
\end{equation}

where $m(v)$ is the exterior mass function, $v$ is the retarded time coordinate, $\tilde r$ is the exterior areal radius, and $\mathcal{Q}$ stands for the total electric charge.

\begin{equation}\label{eq:16}
A(r_{\Sigma},t)\,dt=\left[1-\frac{2m}{\tilde{r}_{\Sigma}} +\frac{\mathcal{Q}_{\Sigma}^{2}}{\tilde{r}_{\Sigma}^{2}}+2\frac{d\tilde{r}_{\Sigma}}{dv} \right]^{1/2} dv,
\end{equation}

and 

\begin{equation}\label{eq:17}
Y(r_{\Sigma},t)=\tilde{r}_{\Sigma}(v).
\end{equation}

These relations ensure continuity of the temporal and areal-radius components across $\Sigma$.

\begin{equation}\label{eq:170}
m(v)_{\Sigma}=\left[\frac{Y}{2}\left(1+\frac{\dot Y^{2}}{A^{2}}+\frac{\mathcal{Q}^{2}}{Y^{2}}-\frac{Y'^{2}}{B^{2}}\right)\right]_{\Sigma}.
\end{equation}

This implies that the total mass energy is not only determined by the geometry of the problem but also the charge.

Furthermore, the junction condition requires that there must be an equilibrium between the radial pressure and heat flux at the surface of the star.

\begin{equation}\label{eq:171}
(p_{r})_{\Sigma}=(qB)_{\Sigma}.
\end{equation}

Hence, the boundary evolution is controlled by a balance between the interior pressure and dissipative energy transport, while the electric charge enters explicitly through the exterior geometry and mass function. Now by equation (\ref{eq:14b}) and (\ref{eq:14d}), the equation (\ref{eq:171}) is formulated as 

\begin{equation}\label{eq:18}
\begin{aligned}
0 ={}&2Y\ddot{Y}+\dot{Y}^{2}-2\left(\frac{A'}{B}+\frac{\dot{A}}{A}\right)Y\dot{Y}+\frac{2A}{B}Y\dot{Y}'
\\[4pt]
&-\frac{2A}{B^{2}}\left(\dot{B}+A'\right)YY'-\frac{A^{2}}{B^{2}}Y'^{2}+A^{2}-\frac{A^{2}\mathcal{Q}_{\Sigma}^{2}}{Y^{2}},
\end{aligned}
\end{equation}

and charge is calculated as

\begin{equation}\label{eq:19}
\begin{aligned}
\mathcal{Q}^{2}={}&\Bigg[2Y\ddot{Y}+\dot{Y}^{2}
-2\left(\frac{A'}{B}+\frac{\dot{A}}{A}\right)Y\dot{Y} +\frac{2A}{B}Y\dot{Y}'
\\[4pt]
&\quad-\frac{2A}{B^{2}}\left(\dot{B}+A'\right)YY'-\frac{A^{2}}{B^{2}}(Y')^{2}+A^{2}\Bigg]
\frac{Y^{2}}{A^{2}}.
\end{aligned}
\end{equation}

The Eq. (\ref{eq:18}) is unsolvable unless further assumptions are made. In order to keep the system simple yet dynamic, we choose to

\begin{equation}\label{eq:20}
A(r)=A,\qquad Y=rR(t), \qquad B=(B(t)),
\end{equation}

where $R(t)$ dictates the time-dependent nature of the area radius. Upon substituting them into Eq. (\ref{eq:18}), we convert the boundary condition into following equation. 

\begin{equation}\label{eq:21}
\begin{aligned}
2r^{2}R\ddot{R} +r^{2}\dot{R}^{2} +\frac{2r}{B}\left(A-rA'\right)R\dot{R}{}
-\frac{A}{B^{2}}&
\\[4pt]
\left(2rA'+2r\dot{B}+A\right)R^{2} +A^{2}-\frac{A^{2}\mathcal{Q}^{2}}{r^{2}R^{2}}=0,
\end{aligned}
\end{equation}

which includes the effect of the electric field on surface development. In the shear-free case, we also have

\begin{equation}\label{eq:22}
A=A(r),\qquad Y=rB(t), \qquad B=B(t),
\end{equation}

so that the radial and time scaling factors change in concert. With such an assumption, Eq.(\ref{eq:21}) simplifies to the nonlinear shear-free differential equation

\begin{equation}\label{eq:23}
2B\ddot{B}+\dot{B}^{2}+\alpha\dot{B}+\beta+\frac{\gamma}{B^{2}}=0,
\end{equation}

where $\alpha =-2A$, $\beta =\frac{-2AA'}{r}$ and $\gamma = \frac{-A^{2} \mathcal{Q}^{2}}{r^{4}}$ are evaluated at the stellar boundary. The parameter $\gamma$ contains the effect of electric charge and distinguishes the present model from the neutral radiating case.

%case 5 https://link.springer.com/article/10.1007/s10714-024-03338-1
A similar charged shear-free evolution equation was investigated by Maharaj and Govinder \cite{S2}. The equation admits an exact solution $\dot{B}=\ddot{B}=0$.

\begin{equation}\label{eq:24}
B(t)= \pm \sqrt{\frac{-\gamma}{\beta}}.
\end{equation}

To study the temporal dynamics, we introduce $x=\dot B$, which converts Eq.(\ref{eq:23}) into the first-order system

\begin{equation}\label{eq:25}
\dot{B} = x, \qquad
2B\dot{x}
=
-x^{2}
-\alpha x
-\beta
-\frac{\gamma}{B^{2}}.
\end{equation}

Here $(B_{\ast},x_{\ast})=\left(\pm\sqrt{-\frac{\gamma}{\beta}},\,0\right)$ represent the charged shear-free configuration.

\section{Solutions}
\label{sec:3}
For deriving explicit solutions of the charge boundary condition, a simple linear functional form of the time-scale factor is assumed,

\begin{equation}\label{eq:26}
R=at,
\end{equation}

where $a$ is a constant. This way, the model remains dynamical but the equation governing it becomes simpler. Substituting Eq. (\ref{eq:26}) into (\ref{eq:21}) into the boundary condition yields

\begin{equation}\label{eq:27}
\begin{aligned}
&0=2ra^{2}At^{2}\dot{B}+2a^{2}\left(r^{2}A'-rA\right)tB
\\[4pt]
&-\left(a^{2}r^{2}+A^{2}+\frac{A^{2}\mathcal{Q}^{2}}{(art)^{2}}
\right)B^{2}+a^{2}\left(A^{2}+2rAA'\right)t^{2}.
\end{aligned}
\end{equation}

If $R=B=at$, the shear is zero. And Eq. (\ref{eq:27}) results

\begin{equation}\label{eq:28}
2A'A + 2arA' - (a^{2}r+\frac{A^{2}\mathcal{Q}^{2}}
{a^{2}r^{3}t^{2}}) = 0.
\end{equation}

This is true when $A = A(r)$. For $R \neq B$,
the conditions at boundary will be more complex. Eq. (\ref{eq:27}) is a Riccati
equation with respect to the variable $B$. It is important to note that
there is no general solution for Riccati equations. However, in our
case, we may simplify things by defining a new variable

\begin{equation} \label{eq:29}
f=\frac{t}{B}.
\end{equation}

Now Eq. (\ref{eq:27}) becomes

\begin{equation}\label{eq:30} 
t\dot{f}=c_{3}f^{2}+c_{2}f+c_{1},
\end{equation}

where

\begin{subequations}\label{eq:31}
\begin{align}
c_{1}&=-\left(\frac{r_{\Sigma}}{2A}+\frac{A}{2a^{2}r_{\Sigma}}+\frac{A(16a^{2}+7r_{\Sigma}^{4})}{2a^{2}r_{\Sigma}^{3}}
\right),
\label{eq:31a}
\\[4pt]
c_{2}&=\frac{r_{\Sigma}A'}{A},\label{eq:30b}\\[4pt]
c_{3}
&=A'+\frac{A}{2r_{\Sigma}}.
\label{eq:31c}
\end{align}
\end{subequations}

The Eq. (\ref{eq:29}) converted the Eq. (\ref{eq:27}) into below form 

\begin{equation}\label{eq:32}
\frac{df}
{c_{3}f^{2}+c_{2}f+c_{1}}
=
\frac{dt}{t}.
\end{equation}

Above equation solution depends on $c_{3}$, $c_{2}$, $c_{1}$.
we take 

\begin{equation}\label{eq:33}
\Delta=4c_{1}c_{3}-c_{2}^{2}.
\end{equation}

To solve Eq. (\ref{eq:32}), we start solving the left side of equation by the following conditions

\begin{align*}
\text{(a)}\quad & \Delta < 0,\\[4pt]
\text{(b)}\quad & \Delta = 0,\quad c_{2}\neq 0,\quad c_{3}\neq 0,\\[4pt]
\text{(c)}\quad & \Delta > 0.
\end{align*}

From the above conditions, to obtain simple results we used the case (b) in which $\Delta = 0$. By using this case the Eq. (\ref{eq:32}) will take the following form  

\begin{equation}\label{eq:34}
-\frac{2}{c_{2}+2c_{3}f}=\ln t + H,
\end{equation}

here $H$ is integration constant. Now separating $f$ from above equation we obtain the following result.

\begin{equation}\label{eq:35}
f=\left(\frac{-2-c_{2}(H+\ln t)}{2c_{3}(H+\ln t)}\right).
\end{equation}

Therefore we have the result of Eq. (\ref{eq:29}) as

\begin{equation}\label{eq:36}
B=\left(\frac{-2c_{3}(H+\ln t)t}{c_{2}(H+\ln t)+2}\right),
\end{equation}

which has an explicit form.

\section{Physical analysis}
\label{sec:4}

In order to test the physical significance of the exact solutions derived in Sect. \ref{sec:3}, we study its properties in detail. The field equations are calculated, and their physical acceptability is verified by the matter distribution and dynamics produced by them. In this regard, solution (\ref{eq:36}) is considered from 3 cases due to the relatively simpler nature of the solution. The energy density, pressure along radial and transverse directions, anisotropy, energy conditions, speed of sound, Herrera's cracking instability criterion, and complexity factor have been determined. All these give a good idea about the physical features of the radiative stellar model. Here the metric function $A(r)$ cannot be determined from this set of solutions, and we take the form

\begin{equation}\label{eq:37}
A(r)=\phi r^{3},
\end{equation}

here, $\phi$ is a constant. Such an assumption makes it possible to retain the analytical nature of the solution while ensuring that a truly dynamical distribution of the physical fields, containing shear, anisotropic stresses, and non-zero heat flow arises. For $r>0$, the matter variables and their stability indicators exhibit well-defined radial profiles. As expected, a singularity develops at $r=0$. A characteristic feature of this particular family of radiating stars is that they develop a central singularity. Other variants of $A(r)$ were also considered. Unfortunately, most of them exhibited some pathological behavior, like singularities or violated stability/ causality conditions. The simple form $R(t)=at$ becomes quite handy since in the absence of shear it allows us to study horizon-free gravitational collapse of a star where compression of the star is compensated by radial heat flow \cite{P1}. In the case of $A(r)=\phi r^{3}$, the four velocity in Eq.(\ref{eq:2}) takes the form

\begin{subequations}\label{eq:38}
\begin{align}
u^{a}&=\left(\frac{1}{\phi r^{3}},\,0,\,0,\,0\right),
\tag{38a}\label{eq:38a}
\\[6pt]
\dot{u}^{a}&=\left(0,\,\frac{3\left(3(H+\ln t)+2\right)^{2}}{
49\phi^{2}r^{5}t^{2}(H+\ln t)^{2}},\,0,\,0\right),
\tag{38b}\label{eq:38b}
\\[6pt]
\Theta&=\frac{1}{\phi r^{3}t}\left(\frac{2}{(\ln t+H)
\left(3(\ln t+H)+2\right)}+3\right),
\tag{38c}\label{eq:38c}
\\[6pt]
\sigma
&=-\frac{2}{3\phi r^{3}t(\ln t+H)\left(3(\ln t+H)+2\right)}.
\tag{38d}\label{eq:38d}
\end{align}
\end{subequations}

For the range $-\infty < t < t_H$ ( $t_H$ represents the formation of the horizon in time), above parameters decrease to zero as the process continues. Such a tendency is a manifestation of the gradual loss of dynamical activity of the system. In subsequent stages, the acceleration decreases faster than the rest of the kinematical variables.

\subsection{Density and pressures}

The system of field equations computed using (\ref{eq:36}), (\ref{eq:37}) and $Y =[rR(t)]$ are 

\begin{subequations}\label{eq:39}
\begin{align}
\rho&=\frac{1}{49r^{6}t^{2}}\Biggl[\frac{1}{\phi^{2}}\left(
\frac{12}{(H+\ln t)^{2}}+\frac{134}{H+\ln t}-\frac{295}{3\ln t+2+3H}+174\right)\nonumber\\
&+\frac{49r^{4}}{a^{2}}\Biggr]+\frac{(16a^{2}+7\phi^{2}r^{4})}{7\phi^{2}r^{8}a^{2}t^{2}},
\tag{39a}\label{eq:39a}
\\[6pt]
p_{r}&=\frac{1}{7r^{6}t^{2}}\left[\frac{\left(3(H+\ln t)+2\right)^{2}}{\phi^{2}(H+\ln t)^{2}}-\frac{7r^{4}}{a^{2}}
-\frac{7}{\phi^{2}}\right]+\frac{(16a^{2}+7\phi^{2}r^{4})}{7\phi^{2}r^{8}a^{2}t^{2}},
\tag{39b}\label{eq:39b}
\\[6pt]
p_{t}&=\Bigg(\frac{4}{49\phi^{2}r^{6}t^{2}(3\ln t+2+3H)^{2}(H+\ln t)^{2}}\nonumber\\
&\quad\times
\Bigg(4-H\left[H\left(93H+90H^{2}+142\right)-73\right]
\nonumber\\
&\quad-\ln t\Bigg[\left((31+40H)9H+284\right)H-73\nonumber\\
&\quad+\ln t\Big(9H(60H+31)+3\ln t(31+30\ln t+120H)+142\Big)\Bigg]\Bigg)\Bigg)\nonumber\\
&\quad-\frac{(16a^{2}+7\phi^{2}r^{4})}{7\phi^{2}r^{8}a^{2}t^{2}}.
\tag{39c}\label{eq:39c}
\end{align}
\end{subequations}

\begin{figure}[!htbp]
\centering
\subfloat[]{\includegraphics[width=70mm]{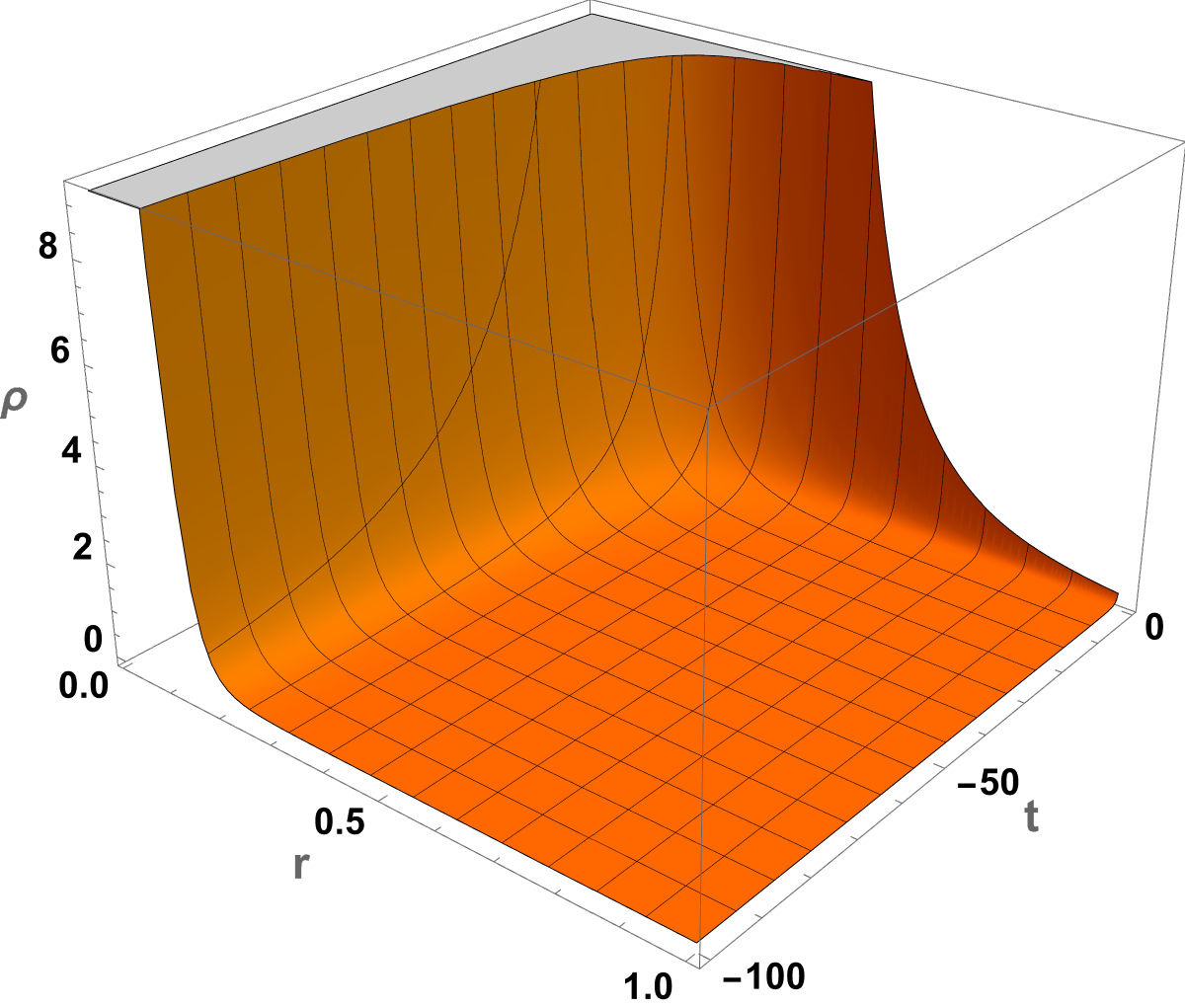}\label{fig:1a}}
\subfloat[]{\includegraphics[width=70mm]{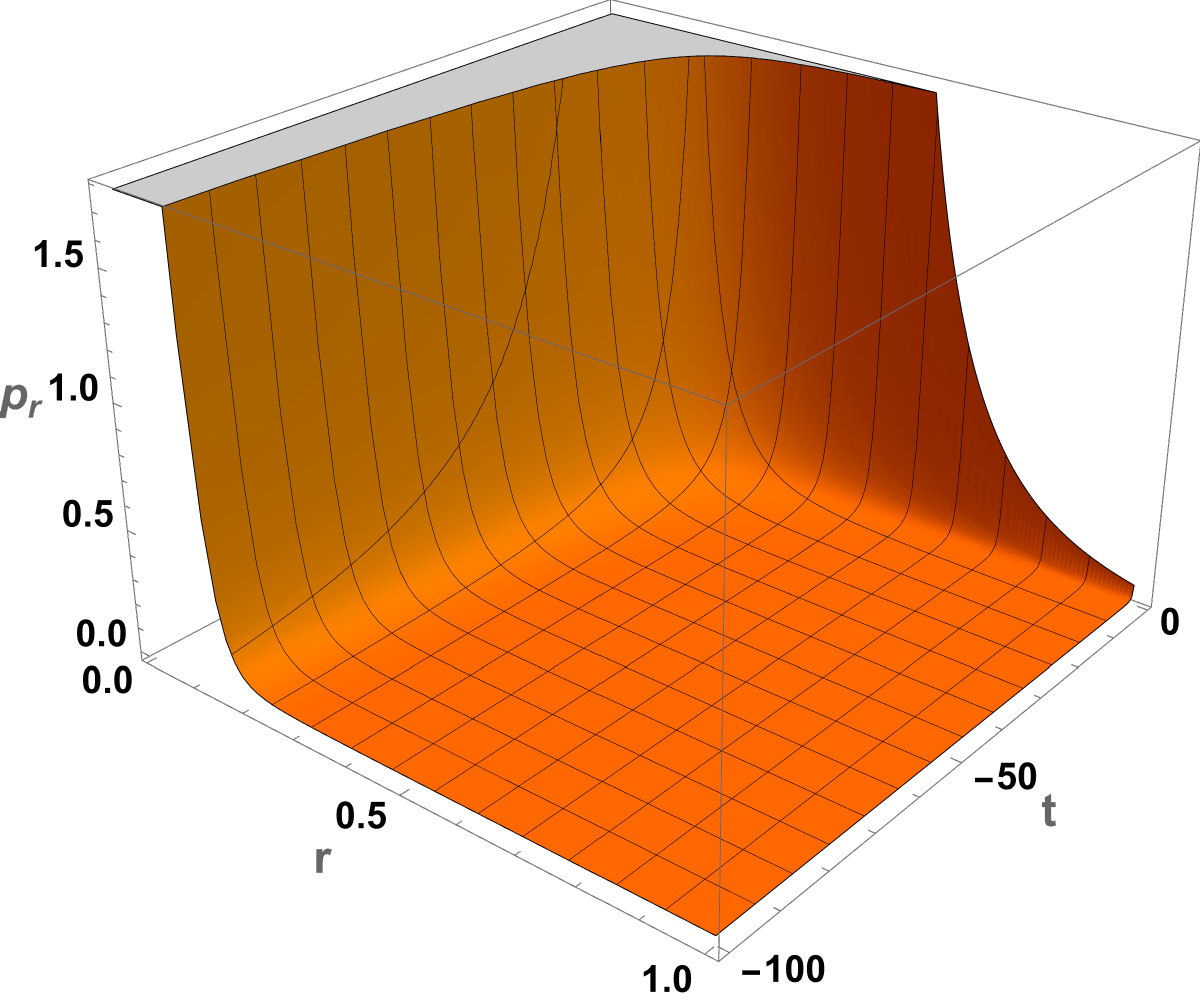}\label{fig:1b}}\\
\subfloat[]{\includegraphics[width=70mm]{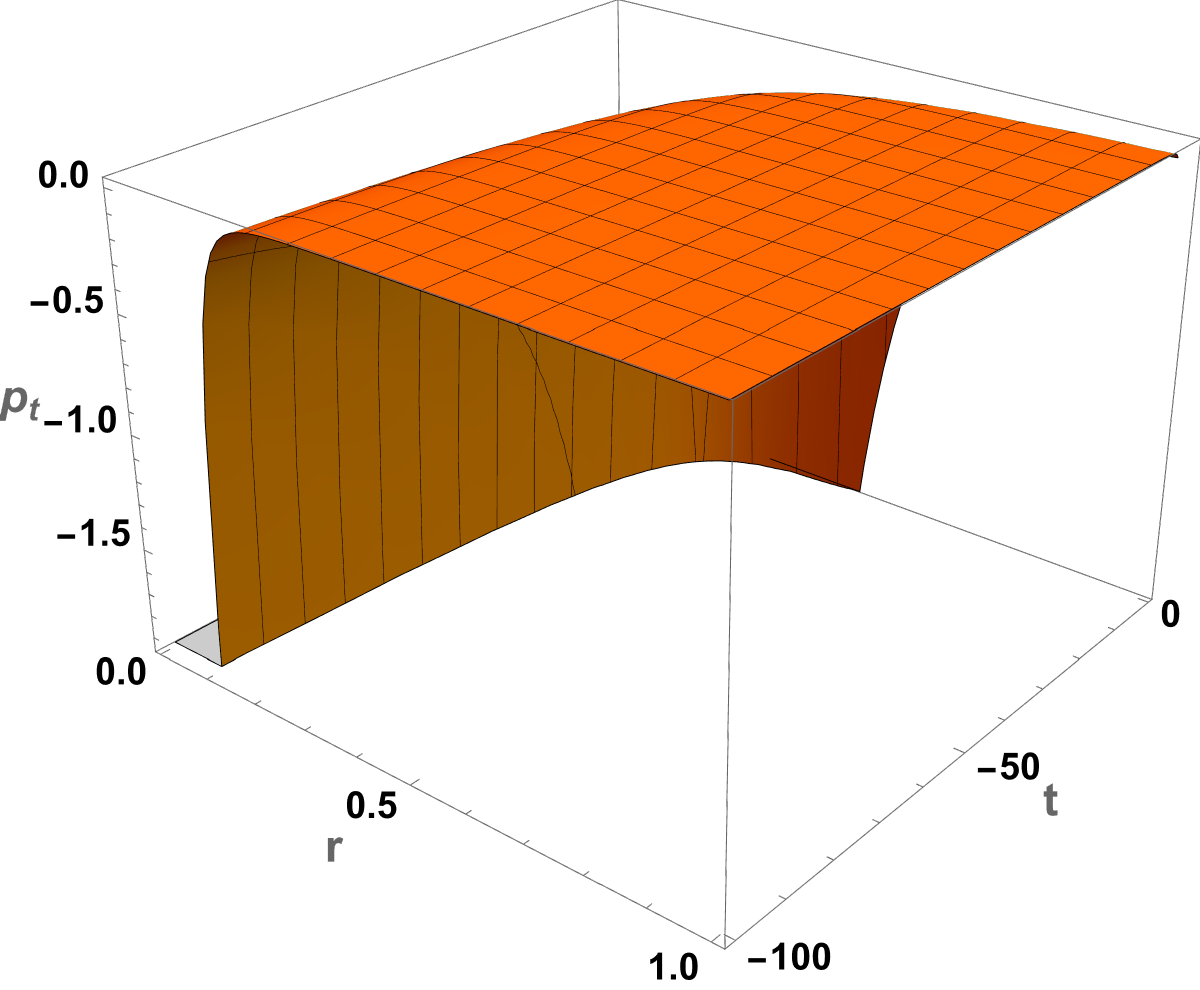}\label{fig:1c}}
\hspace{200mm}
\caption{Evolution of (a) density $\rho$ in $km^{-2}$, (b) radial pressure $p_{r}$ in $km^{-2}$, and (c) tangential pressure $p_{t}$ in in $km^{-2}$, with radial coordinate $r$ and time $t$ for $H=1$, $\phi=1$, and $a=1$.}
\label{fig:1}
\end{figure}

 Density $\rho$ is plotted against the variables r and t in Fig. (\ref{fig:1a}). It always remains positive and attains the maximum values within the central core of the star. Density goes on decreasing with increasing value of the radial variable and attains low values at the edge.

 The variation of radial pressure $p_r$ is shown in Fig. (\ref{fig:1b}). Like the density, $p_r$ is positive and significant in the inner parts of the star. It smoothly decreases towards the outer surface of the sphere and remains smaller than the densities in the same region. This shows that there is consistent radial pressure during the process of collapse.

Fig. (\ref{fig:1c}) represents the variation of tangential pressure $p_t$. Unlike the density and radial pressure, tangential pressure is negative throughout the region and its value is larger on the inner side but decreases outward. The difference in the values of $p_r$ and $p_t$ is clear evidence of the anisotropy of the charged sphere.

\FloatBarrier

\subsection{Heat flux and charge}

The heat flux and charge equations are simplified by using (\ref{eq:36}), (\ref{eq:37}) and $Y =rR(t)$ as

\begin{equation}\label{eq:40}
Q_{h}=-\frac{2(3\ln t+3H+2)}{49\phi^{3}r^{8}t^{3}(\ln t+H)^{3}
}(3\ln t\left(3\ln t+6H+2\right)+9H^{2}+6H+2).
\end{equation}

\begin{figure}[!htbp]
\centering
\subfloat[]{\includegraphics[width=70mm]{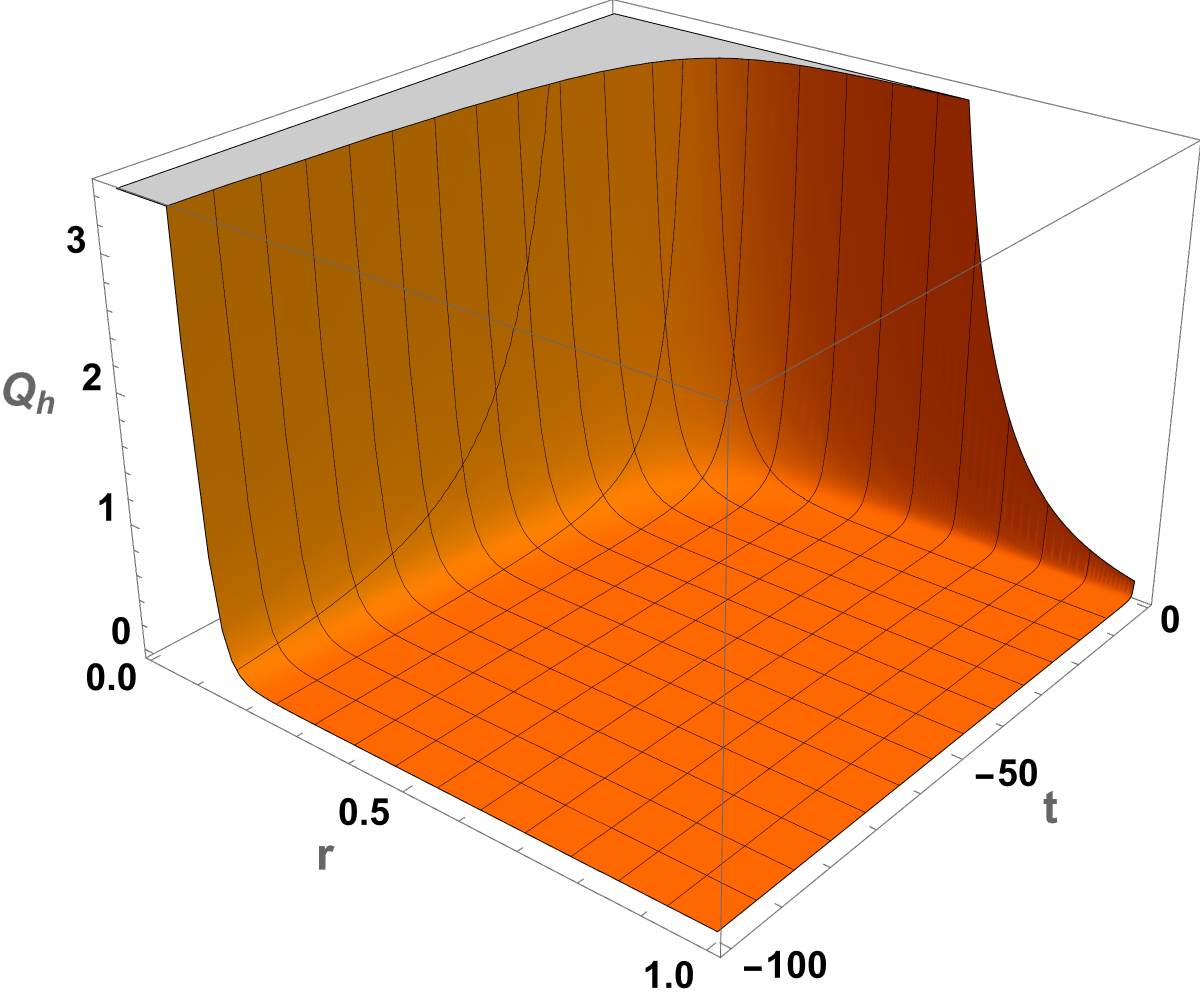}\label{fig:2a}}
\subfloat[]{\includegraphics[width=70mm]{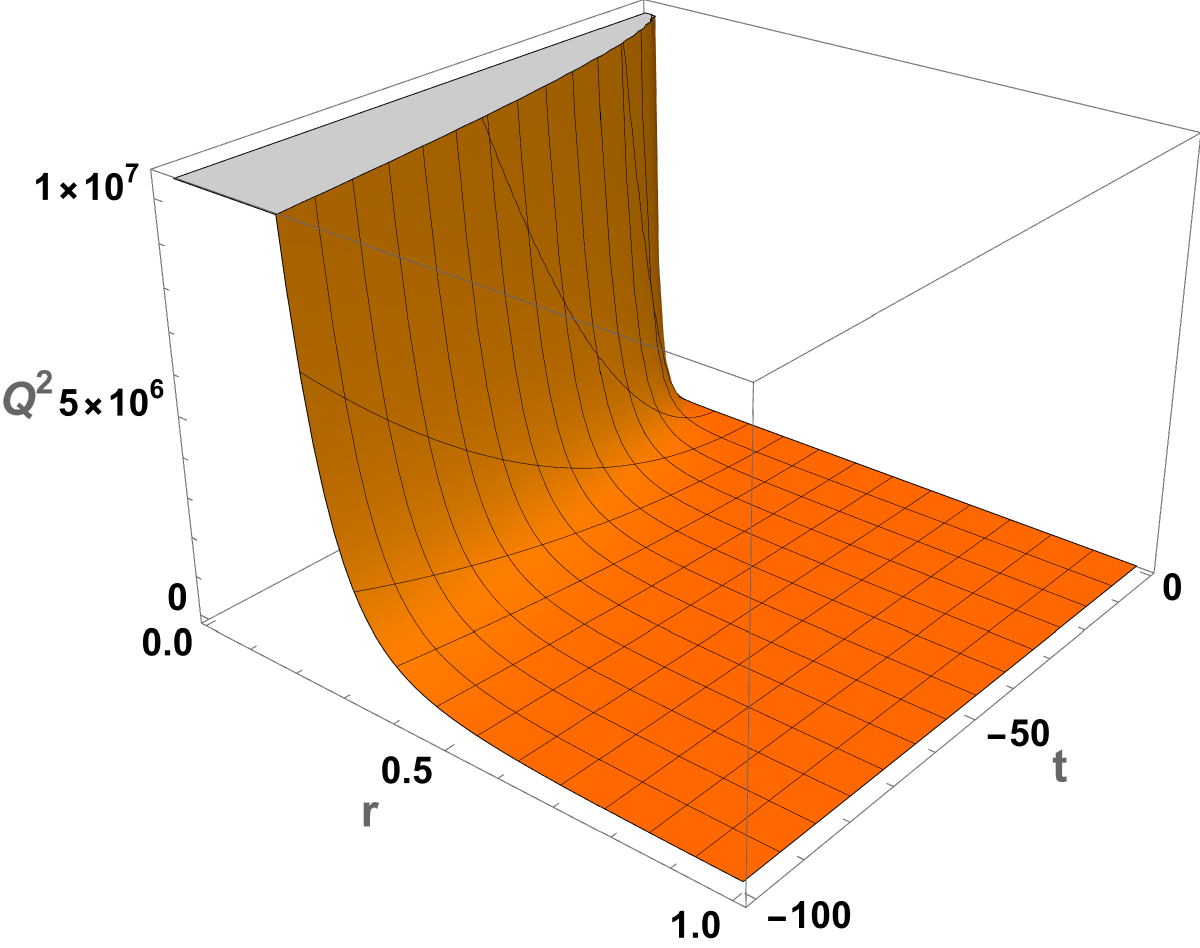}\label{fig:2b}}
\hspace{200mm}
\caption{Profiles of (a) radial heat flux $Q_{h}$ in $km^{-2}$ and (b) squared charge function $Q^{2}$ in $km^{2}$ as functions of $r$ and $t$ for $H=\phi=a=1$.}
\label{fig:2}
\end{figure}

The heat flux $Q_h$ (see Fig. (\ref{fig:2a})) stays positive and peaks within the inner regions of the star. The value of the heat flux diminishes sharply with an increase in $r$, which means that energy transfer is higher in the deeper stellar interior and weaker in the proximity of its surface. This function varies with time due to the dissipative evolution of the system.

\begin{equation}\label{eq:41}
Q^{2}=\frac{a^{2}t^{2}(16a^{2}+7\phi^{2}r^{4})}{7\phi^{2}r^{4}}.
\end{equation}

In Fig. (\ref{fig:2b}) we see how the function $\mathcal{Q}^{2}$ behaves. The charge term is localized strongly within the inner regions and diminishes sharply when $r$ becomes larger. This function also depends on time and diminishes with the progression of the collapse process. It means that the electromagnetic term dominates in the compact inner regions of the star.

\FloatBarrier

\subsection{Energy conditions}

The energy conditions are used to check the physical validity of the matter distribution in a relativistic stellar model. While there may be many solutions mathematically possible from the Einstein field equations, not all of them will satisfy some energy conditions. In the exact solution of this paper, energy conditions will help us in determining the behaviour of energy density, anisotropic pressure, heat flux, and electromagnetic field during the collapse process. The energy-momentum tensor in Eq. (\ref{eq:3}) is anisotropic. Therefore, the eigenvalues are taken to be real and can be used to write down the energy conditions, with $\lambda_{0}$ being the timelike eigenvalue and $\lambda_{i}$ being the spacelike eigenvalues, with $i\in{1,2,3}$.

\text{(a) Null energy conditions}

\begin{subequations}\label{eq:42}
\begin{equation}
\lambda_{i}-\lambda_{0}\geq 0,
\qquad
i\in\{1,2,3\}.
\label{eq:42a}
\end{equation}

\text{(b) Weak energy conditions}

\begin{equation}
\lambda_{i}-\lambda_{0}\geq 0,
\qquad
(-\lambda_{0})\geq 0,
\qquad
i\in\{1,2,3\}.
\label{eq:42b}
\end{equation}

\text{(c) Dominant energy conditions}

\begin{equation}
\lambda_{0}\leq(\lambda_{i})\leq-\lambda_{0},
\qquad
(-\lambda_{0})\geq 0,
\qquad
i\in\{1,2,3\}.
\label{eq:42c}
\end{equation}

\text{(d) Strong energy conditions}

\begin{equation}
-\lambda_{0}
+\sum_{i=1}^{3}(\lambda_{i})\geq 0,
\qquad
\lambda_{i}-(\lambda_{0})\geq 0,
\qquad
i\in\{1,2,3\}.
\label{eq:42d}
\end{equation}
\end{subequations}

The Eq. (\ref{eq:3}) with charge admits eigenvalues $\lambda$ determined by solving the equation

\begin{equation}\label{eq:43}
\left|(T_{ab}+E_{ab})-\lambda g_{ab}\right|=0.
\end{equation}

By solving Eq. (\ref{eq:43}) with Eq. (\ref{eq:1}) and Eq. (\ref{eq:3}), we get

\begin{equation}\label{eq:44}
\resizebox{0.92\textwidth}{!}{$
\left|
\begin{array}{cccc}
A^{2}(\lambda+\rho+\frac{\mathcal{Q}^{2}}{8\pi Y^{4}}) & -AB^{2}q & 0 & 0 \\[4pt]
-AB^{2}q & B^{2}(p_{r}-\lambda-\frac{\mathcal{Q}^{2}}{8\pi Y^{4}}) & 0 & 0 \\[4pt]
0 & 0 & Y^{2}(p_{t}+\lambda+\frac{\mathcal{Q}^{2}}{8\pi Y^{4}}) & 0 \\[4pt]
0 & 0 & 0 & Y^{2}\sin^{2}\theta(p_{t}+\lambda+\frac{\mathcal{Q}^{2}}{8\pi Y^{4}})
\end{array}
\right|
=0.
$}
\end{equation}

Finding the determinant of the LHS makes the equation
above reduce to

\begin{equation}\label{eq:45}
\begin{aligned}
&AB^{2}\left[Y^{2}
\left(
p_{t}+\lambda+\frac{\mathcal{Q}^{2}}{8\pi Y^{4}} 
\right)\right]\left[Y^{2}\sin^{2}\theta
\left(
p_{t}
+\lambda +\frac{\mathcal{Q}^{2}}{8\pi Y^{4}} 
\right)\right]
\\[4pt]
&\times \left[A \lambda\left(
-\frac{B^{2}q^{2}}{\lambda}
-\lambda
-\rho
+ p_{r}
- \frac{\mathcal{Q}^{2}}{4\pi Y^{4}} 
\right)+A\left(\rho+\frac{\mathcal{Q}^{2}}{8\pi Y^{4}} \right)\left(p_{r}-\frac{\mathcal{Q}^{2}}{8\pi Y^{4}} \right)\right]
=0.
\end{aligned}
\end{equation}

Simplifying above equation gives one form like 

\begin{equation}\label{eq:46}
\lambda^{2}+(\rho-p_{r}+\frac{\mathcal{Q}^{2}}{8\pi Y^{4}} )\lambda+B^{2}q^{2}-\left(\rho+\frac{\mathcal{Q}^{2}}{8\pi Y^{4}} \right) \left(p_{r} -\frac{\mathcal{Q}^{2}}{8\pi Y^{4}} \right).    
\end{equation}

Further, Eq. (\ref{eq:46}) results

\begin{equation}\label{eq:47}
\lambda_{0}=\frac{1}{2}\left[p_{r}-D-\rho-\frac{\mathcal{Q}^{2}}{8\pi Y^{4}}\right],
\qquad \lambda_{1}=-\frac{1}{2}\left[\rho-p_{r}-D +\frac{\mathcal{Q}^{2}}{8\pi Y^{4}}\right].
\end{equation}

Where

\begin{equation}\label{eq:48}
D^{2}=\left(p_{r}+\rho\right)^{2}-4Q_{h}^{2}\geq 0,
\qquad |p_{r}+\rho|-2|Q_{h}|\geq 0.
\end{equation}

The second solution arising from Eq. (\ref{eq:45}) is below which has the root $\lambda_{2}$.

\begin{equation}\label{eq:49}
Y^{2}\sin^{2}\theta\,\left(p_{t}+\lambda+\frac{\mathcal{Q}^{2}}{8\pi Y^{4}}\right)=0, \qquad \lambda_{2}=-\left(p_{t}+\frac{\mathcal{Q}^{2}}{8\pi Y^{4}}\right).
\end{equation}

Third solution from Eq. (\ref{eq:45}) with its calculated root is written as

\begin{equation}\label{eq:50}
AB^{2}\left(p_{t}+\lambda +\frac{\mathcal{Q}^{2}}{8\pi Y^{4}}\right),\qquad \lambda_{3}=\lambda_{2}=-\left(p_{t}+\frac{\mathcal{Q}^{2}}{8\pi Y^{4}}\right).
\end{equation}

In order for the eigenvalues to be real-valued, the discriminant should adhere to the condition $D \geq 0$ in the relevant spacetime domain. In light of this condition, one is able to formulate the null, weak, dominant, and strong energy conditions through the eigenvalues of the energy-momentum tensor.

\subsubsection{Null energy conditions (NEC)}

With Eq. (\ref{eq:42a}), (\ref{eq:47}), (\ref{eq:49}), and $i = 1$ gives the following condition

\begin{subequations}\label{eq:51}
\begin{equation}\label{eq:51a}
D\geq 0,
\end{equation}

thus from Eq. (\ref{eq:48}), Eq. (\ref{eq:42a}), $i = 2$, (\ref{eq:47})
and (\ref{eq:49}), the energy condition becomes

\begin{equation}\label{eq:51b}
\rho-2p_{t}+D-p_{r}\geq 0.
\end{equation}
\end{subequations}

\subsubsection{Weak energy conditions (WEC)}

Substituting the eigenvalues from Eq. (\ref{eq:47}), and (\ref{eq:49}) into Eq. (\ref{eq:42b})
yields the weak energy conditions

\begin{subequations}\label{eq:52}
\begin{equation}\label{eq:52a}
\rho-p_{r}+D+\frac{\mathcal{Q}^{2}}{4\pi Y^{4}}\geq 0,
\end{equation}
\begin{equation}\label{eq:52b}
D\geq 0,
\end{equation}
\begin{equation}\label{eq:52c}
\rho-p_{r}+D-\frac{\mathcal{Q}^{2}}{4\pi Y^{4}}\geq 0.
\end{equation}
\end{subequations}

The weak energy conditions obtained above also include the null energy constraints presented in Eqs. Eqs. (\ref{eq:51a}) and
(\ref{eq:51b}).

\subsubsection{Dominant Energy Conditions (DEC)}

Using the eigenvalues from Eqs. (\ref{eq:47}), and (\ref{eq:49}) into the Eq. (\ref{eq:42c}) , 
the dominant energy becomes

\begin{subequations}\label{eq:53}
\begin{equation}\label{eq:53a}
\rho+D-p_{r}+\frac{\mathcal{Q}^{2}}{4\pi Y^{4}}\geq 0,
\end{equation}
\begin{equation}\label{eq:53b}
D\geq 0
\end{equation}
\begin{equation}\label{eq:53c}
\rho-p_{r}+\frac{\mathcal{Q}^{2}}{4\pi Y^{4}}\geq 0,
\end{equation}
\begin{equation}\label{eq:53d}
p_{t}\leq \frac{1}{2} \left(\rho+D-p_{r}\right),
\end{equation}
\begin{equation}\label{eq:53e}
p_{t}\leq \frac{1}{2} \left(\rho +D-p_{r}\right)+\frac{\mathcal{Q}^{2}}{4\pi Y^{4}}.
\end{equation}
\end{subequations}

It should be noted that the first two inequalities (\ref{eq:53a}) and (\ref{eq:53b}) hold true for the weak energy conditions (\ref{eq:52a}) and (\ref{eq:52b}), whereas
the rest of the inequalities (\ref{eq:53c})–(\ref{eq:53e}) represent additional dominant conditions.

\subsubsection{Strong energy conditions (SEC)}

By Eqs. (\ref{eq:47}), (\ref{eq:49}), (\ref{eq:45}) and (\ref{eq:42d}) the conditions becomes

\begin{subequations}\label{eq:54}
\begin{equation}\label{eq:54a}
D-2p_{t}-\frac{\mathcal{Q}^{2}}{4\pi Y^{4}}\geq 0,
\end{equation}
\begin{equation}\label{eq:54b}
D\geq 0,
\end{equation}
\begin{equation}\label{eq:54c}
p_{t}\leq \frac{1}{2} \left(\rho+D-p_{r}\right).
\end{equation}
\end{subequations}

Here, the inequality (\ref{eq:54b}) is followed by the
NEC, DEC, and WEC, and the inequality (\ref{eq:54c}) is implied
by (\ref{eq:53d}). The Eq. (\ref{eq:54a}) gives the
newly introduced SEC.\\

We briefly present the different energy inequalities as

\begin{subequations}\label{eq:55}
\begin{align}
E_{1}: \quad
&(p_{r}+\rho)^{2}-4Q_{h}^{2}\geq 0
\qquad
(\mathrm{NEC/DEC/WEC/SEC}),
\label{eq:55a}
\\[4pt]
E_{2}: \quad
&-p_{r}+D+\rho+\frac{\mathcal{Q}^{2}}{4\pi Y^{4}}\geq 0
\qquad
(\mathrm{WEC/DEC}),
\label{eq:55b}
\\[4pt]
E_{3}: \quad
&-2p_{t}+D+\rho-p_{r}\geq 0
\qquad
(\mathrm{NEC/WEC}),
\label{eq:55c}
\\[4pt]
E_{4}: \quad
&-p_{r}+\frac{\mathcal{Q}^{2}}{4\pi Y^{4}}+\rho\geq 0
\qquad
(\mathrm{DEC}),
\label{eq:55d}
\\[4pt]
E_{5}: \quad
&p_{t}\geq
\frac{\left(
-p_{r}+D+\rho
\right)}{2}
\qquad
(\mathrm{DEC}),
\label{eq:55e}
\\[4pt]
E_{6}: \quad
&-p_{t}\geq
\frac{\left(
-p_{r}+D+\rho
\right)}{2}
+\frac{\mathcal{Q}^{2}}{4\pi Y^{4}}
\qquad
(\mathrm{DEC}),
\label{eq:55f}
\\[4pt]
E_{7}: \quad
&D-2p_{t}-\frac{\mathcal{Q}^{2}}{4\pi Y^{4}}\geq 0
\qquad
(\mathrm{SEC}).
\label{eq:55g}
\end{align}
\end{subequations}

In order to check the physical acceptability of the current model, we need to consider the conditions $E_{1}$ to $E_{3}$. These conditions give us the most significant physical restrictions in terms of reality of eigenvalues, density vs radial pressure, as well as the effect of $p_{t}$, charge and heat conduction. The other conditions, $E_{4}$ to $E_{7}$, are derived from the same physical restrictions and can be checked analytically when $E_{1}$ to $E_{3}$ are fulfilled.

\begin{figure}[!htbp]
\centering
\subfloat[]{\includegraphics[width=70mm]{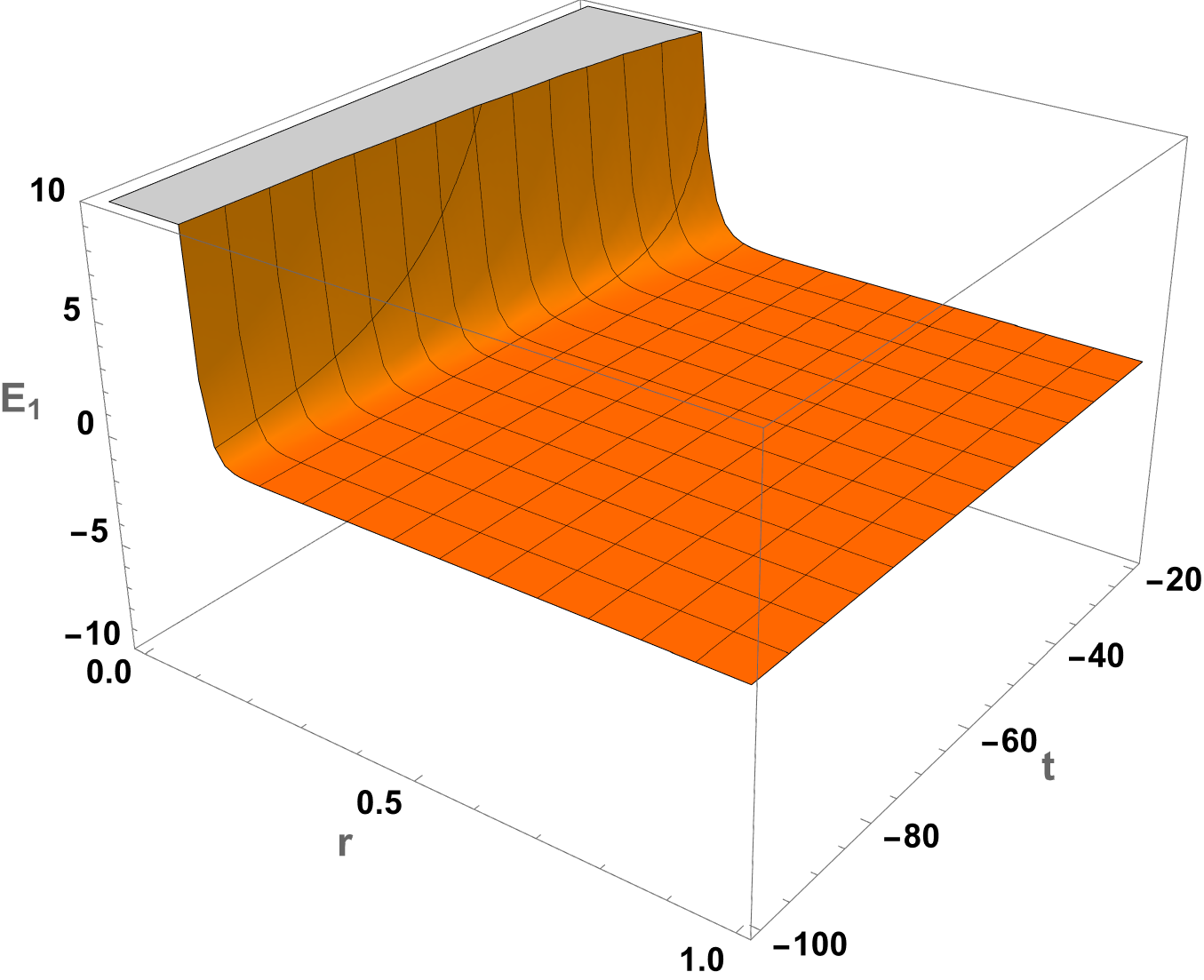}\label{fig:3a}}
\subfloat[]{\includegraphics[width=70mm]{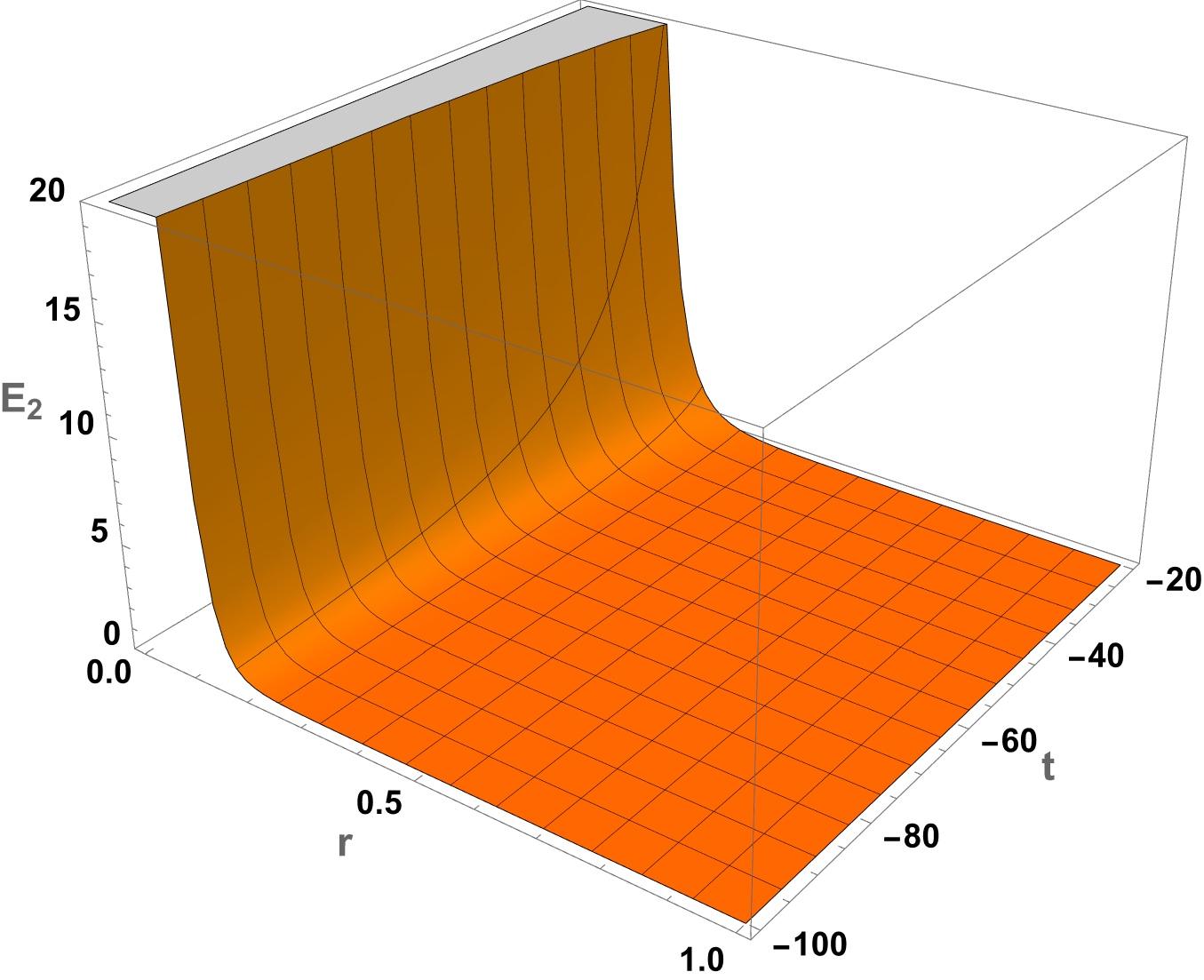}\label{fig:3b}}\\
\subfloat[]{\includegraphics[width=70mm]{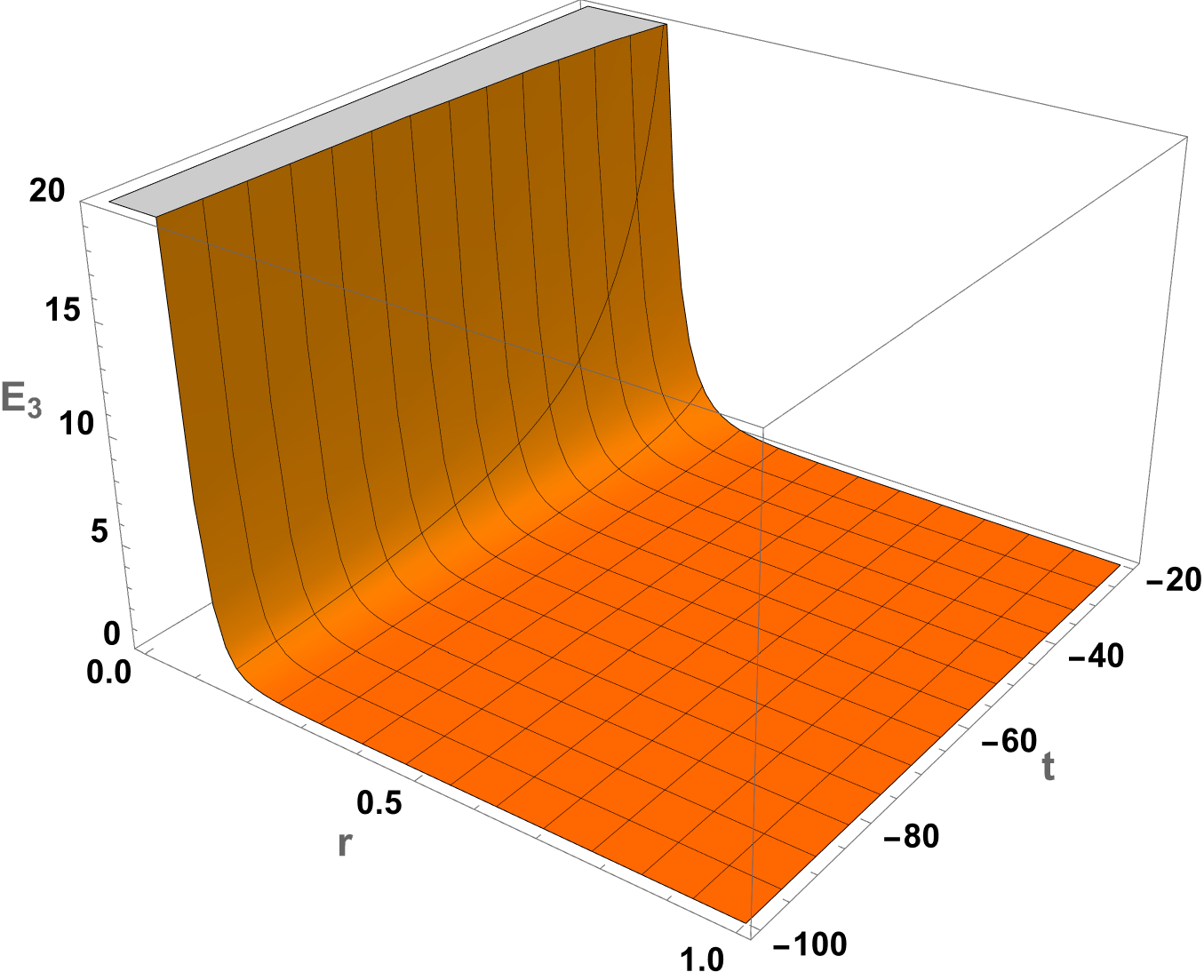}\label{fig:3c}}
\hspace{200mm}
\caption{Profiles of the energy-condition functions (a) $E_{1}(r,t)=(p_{r}+\rho)^{2}-4Q_{h}^{2}$, (b) $E_{2}(r,t)=-p_{r}+D+\rho+\frac{\mathcal{Q}^{2}}{4\pi Y^{4}}$, and (c) $E_{3}(r,t)=-2p_{t}+D+\rho-p_{r}$ as functions of $r$ and $t$ for $H=\phi=a=1$.}
\label{fig:3}
\end{figure}
\FloatBarrier

Fig. (\ref{fig:3a}) presents the behavior of $E_{1}(r,t)=(p_{r}+\rho)^{2}-4Q_{h}^{2}$, which has both positive and negative values throughout the spacetime interval of interest. In the neighborhood of the inner region, $E_{1}$ stays positive, implying that the sum of density and radial pressure overwhelms the heat flux effect and ensures the reality of eigenvalues. However, as the radial variable increases, $E_{1}$ becomes smaller and changes to negative values, revealing that the dissipation effect becomes relatively more pronounced. Hence, the condition $E_{1}\geq0$ holds good only in outer layers and later times part of the configuration.\\
Fig. (\ref{fig:3b}) shows the evolution of $E_{2}(r,t)=\rho-p_{r}+D+\mathcal{Q}^{2}/(4\pi Y^{4})$. The expression remains positive for $r$ and $t$, $r\in[0,1]$ and $t\in[-100,-20]$. This confirms the satisfaction of this energy condition through the process. \\
Fig. (\ref{fig:3c}) illustrates the characteristics of $E_{3}(r,t)=-2p_{t}+D+\rho-p_{r}$, which stays positive at all times within the star. This ensures that the combined influence of radial and tangential pressure, density, and dissipated energy is taken care of. The positivity of this expression reveals that the anisotropic pressure does not violate the energy criteria during the collapse phase of the star.

\subsection{Causality, sound speed, and Herrera’s cracking
condition}

The stability of anisotropic stars is generally considered in terms of the causality restrictions on the sound speed of radius and tangential directions expressed by

\begin{equation}\label{eq:56}
0 \leq v_{sr}^{2}=\frac{dp_{r}}{d\rho}\leq 1,\qquad\text{and}
\qquad0 \leq v_{st}^{2}=\frac{dp_{t}}{d\rho}\leq 1.
\end{equation}

\begin{figure}[!htbp]
\centering
\includegraphics[width=140mm]{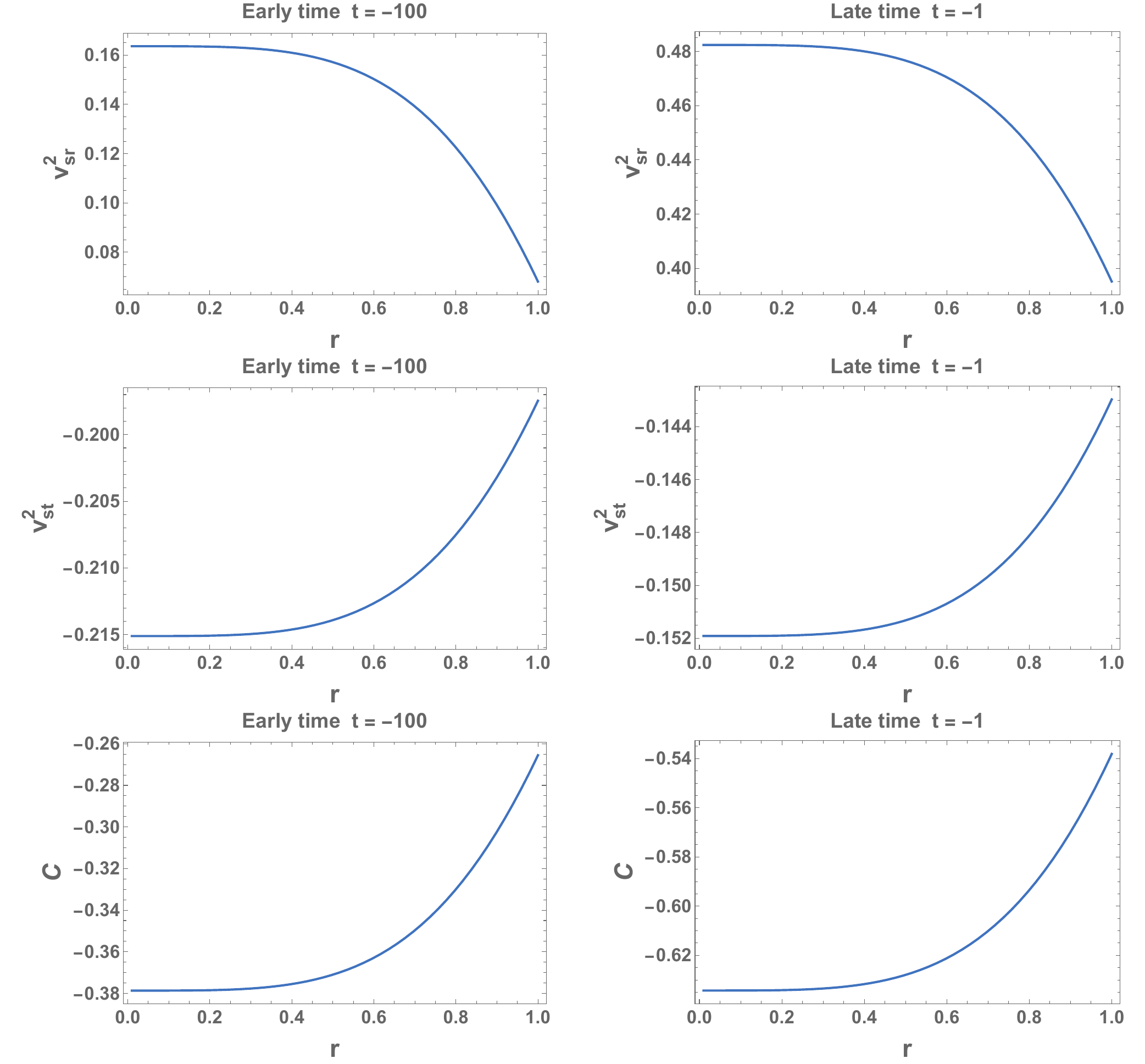}
\hspace{200mm}
\caption{Radial sound speed $v_{sr}^{2}$ (top), tangential sound speed $v_{st}^{2}$ (middle), and cracking function $C=v_{st}^{2}-v_{sr}^{2}$ (bottom) versus the radial coordinate $r\in [0,1]$ at early time $t=-100$ and late time $t=-1$, for $H=\phi=a=1$.}
\label{fig:4}
\end{figure}

The sound speeds are analyzed in terms of their radial and tangential values (at ($t$= -100, and -1), to check the causality conditions. The resulting radial $v_{sr}^{2}$ and tangential $v_{st}^{2}$ sound-speed profiles are shown in Fig. (\ref{fig:4}). Another test for the local stability concerns the cracking condition by Herrera \cite{v1,v2}, where one compares the reaction of the radial and tangential perturbation through their sound speed difference.

\begin{equation}\label{eq:57}
C(r,t)=v_{st}^{2}-v_{sr}^{2}.
\end{equation}

This indicator divides the internal region of a star into regions of Potential stability ($C<0$), and Potential instability ($C>0$) \cite{v3}. In the current case of a radiation field, ($C(r,t)$) is obtained for ($r\in [0,1]$) at various instants ($t$= -100, and -1). The cracking patterns are shown in Fig. (\ref{fig:4}).

\FloatBarrier

It is evident from Fig. (\ref{fig:4}) that $v_{sr}^{2}$ lies in the causal range $0\leq v_{sr}^{2}\leq1$ both in early $(t=-100)$ and late $(t=-1)$ times. $v_{sr}^{2}$ decays monotonically from interior to the surface, but in later times. Clearly, the later time values are higher and hence imply a stiffer response of the fluid to the evolution of the charged system.\\
The tangential velocity squared ($v_{st}^{2}$) for the charged collapse is plotted in Fig. (\ref{fig:4}). Unlike the radial one, the tangential velocity is found to be negative in the entire star region at $t = -100$ and $t = -1$ and hence does not fulfill the causality condition in the tangential direction. The lower value obtained at $t=-1$ signifies that the tangential instability becomes smaller but does not get vanished. \\
Cracking function ($C(r,t)=v_{st}^{2}-v_{sr}^{2}$) is illustrated in Fig. (\ref{fig:4}) and found to remain negative for all times both early and late in the stellar interior. As $C$ is negative for all values of the radial coordinate, the charge distribution satisfies the stable part of the Herrera’s cracking condition.

\section{Anisotropy and Complexity factor}
\label{sec:5}
%https://link.springer.com/article/10.1140/epjc/s10052-025-14147-4
The complexity factor is helpful in defining the internal structure of a stellar model through the combination of effects due to anisotropic pressure, density, and dissipative forces. The complexity factor hence becomes a convenient quantity for measuring the deviation of the fluid away from homogeneity. The complexity factor for a shear, charge, and radiating stellar model is thus defined according to Herrera \cite{Y1} as

\begin{equation}\label{eq:58}
\begin{aligned}
Y_{TF}={}&-8\pi\Pi+\frac{4\pi}{Y^{3}}\int_{0}^{r}Y^{3}\left(\rho'
-\frac{3qBU}{Y}\right)\,dr
-\frac{3}{Y^{3}}\int_{0}^{r}\frac{\mathcal{Q}\mathcal{Q}'}{Y}\,dr\\[4pt]
&+\frac{4\mathcal{Q}^{2}}{Y^{4}},
\end{aligned}
\end{equation}

where anisotropy is defined as

\begin{equation}\label{eq:59}
\Pi = p_{t}-p_{r}.
\end{equation}

The velocity field $U$ describes the speed of change of the areal radius in the direction of the fluid flow. The sign of the velocity is indicative of the dynamic nature of the system, where ($U<0$) implies contraction and ($U>0$) implies expansion \cite{Y1}.

\begin{equation}\label{eq:60}
U= \frac{\dot {Y}}{A}.
\end{equation}

The Eq. (\ref{eq:58}) is divided into smalls terms

\begin{subequations}\label{eq:61}
\begin{align}
Z_{1}&=
-8\pi\Pi,\label{eq:61a}
\\[4pt]
Z_{2}&=\frac{4\pi}{Y^{3}}\int_{0}^{r}Y^{3}\rho'\,dr,
\label{eq:61b}
\\[4pt]
Z_{3}&=-\frac{4\pi}{Y^{3}}\int_{0}^{r}3qBUY^{2}\,dr,
\label{eq:61c}
\\[4pt]
Z_{4}&=-\frac{3}{Y^{3}}\int_{0}^{r}\frac{\mathcal{Q}\mathcal{Q}'}{Y}\,dr+\frac{4\mathcal{Q}^{2}}{Y^{4}}.
\label{eq:61d}
\end{align}
\end{subequations}

From Eq. (\ref{eq:58}), the $\Pi$ is counterbalanced by the $\rho$ inhomogeneities. 

\begin{figure}[!htbp]
\centering
\subfloat[]{\includegraphics[width=70mm]{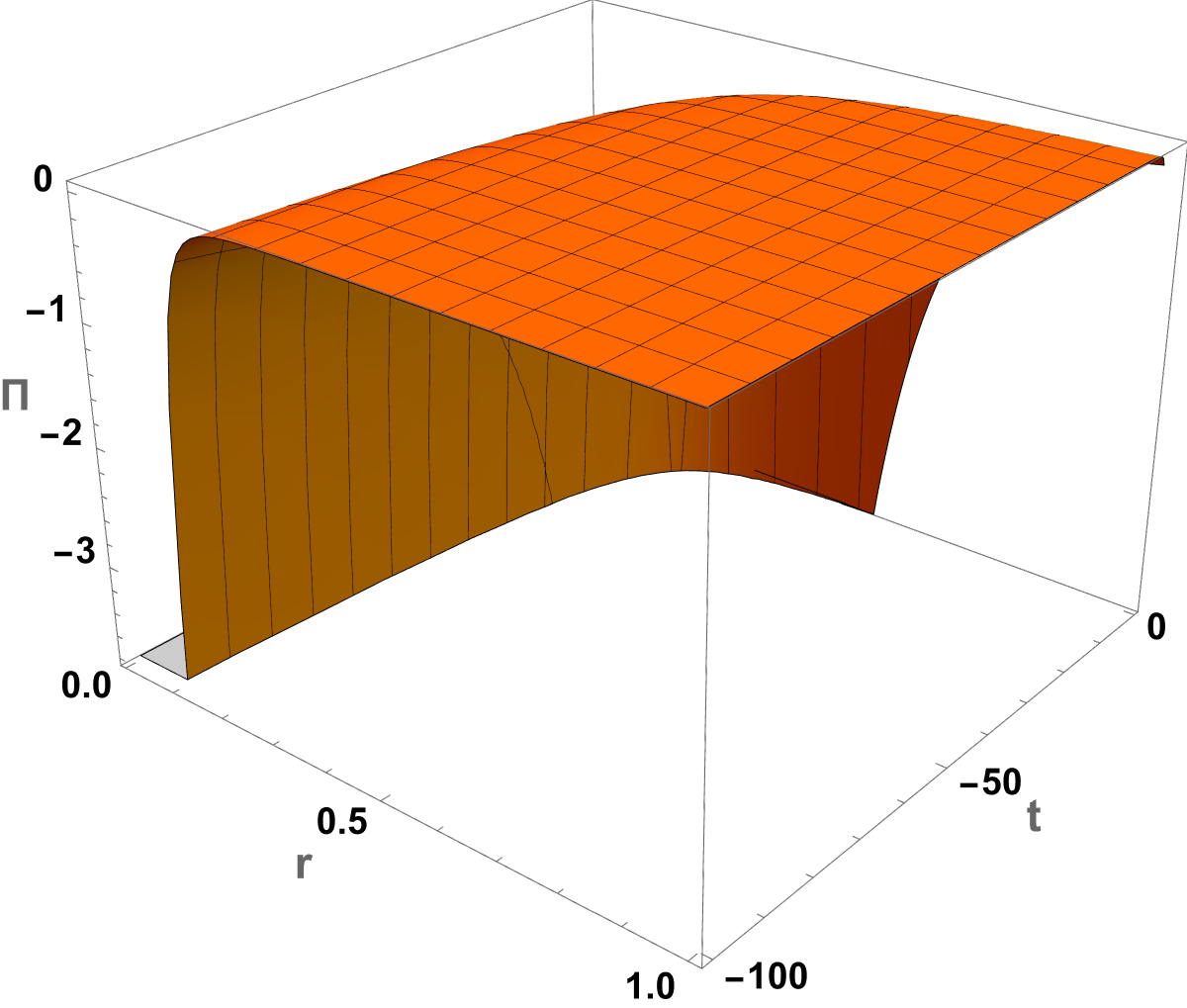}\label{fig:5a}}
\subfloat[]{\includegraphics[width=80mm]{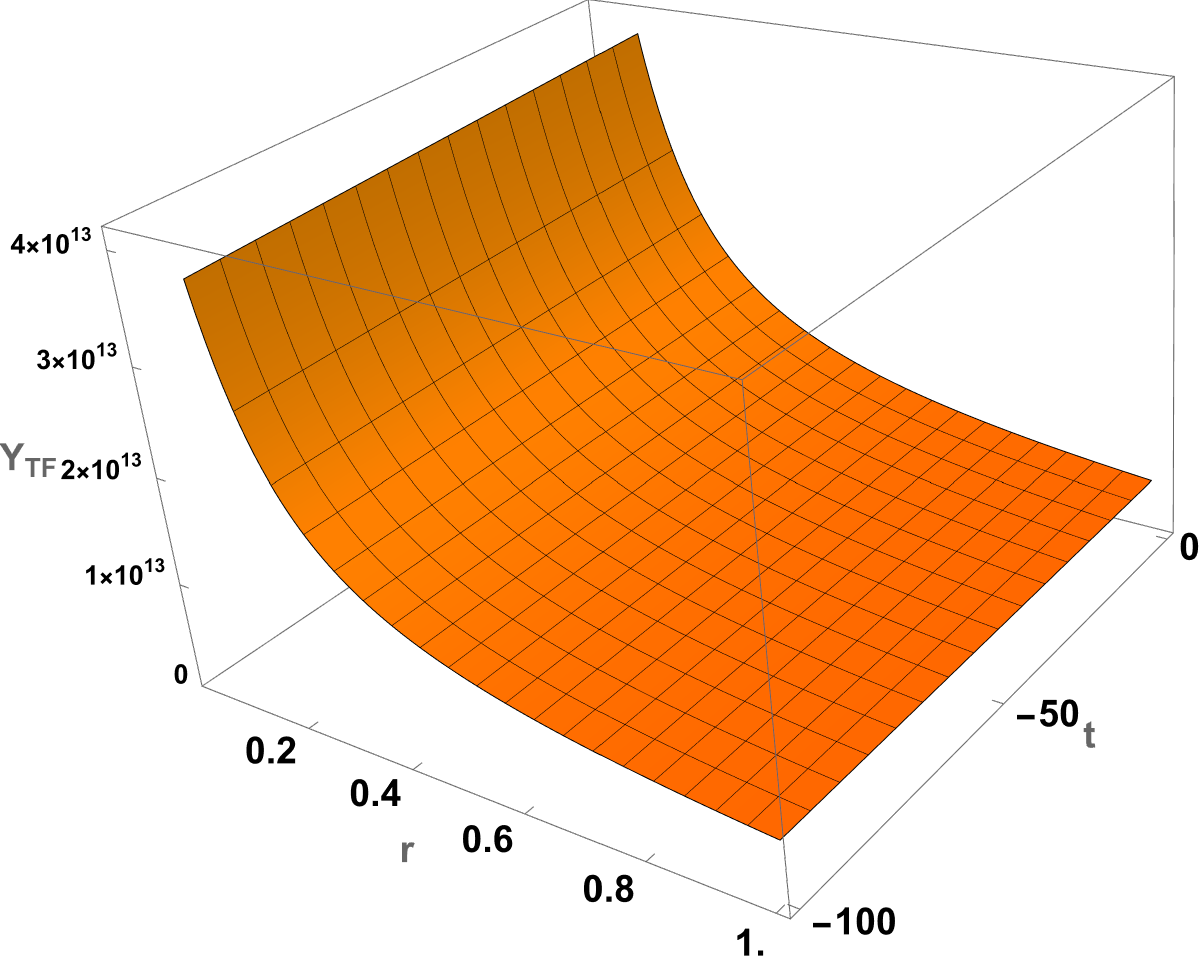}\label{fig:5b}}
\hspace{200mm}
\caption{Profiles of (a) anisotropy $\Pi$ and (b) the complexity factor $Y_{TF}$ inside the star, for $\phi= H= a=1$.}
\label{fig:5}
\end{figure}
\FloatBarrier

\begin{figure}[!htbp]
\centering
\subfloat[]{\includegraphics[width=100mm]{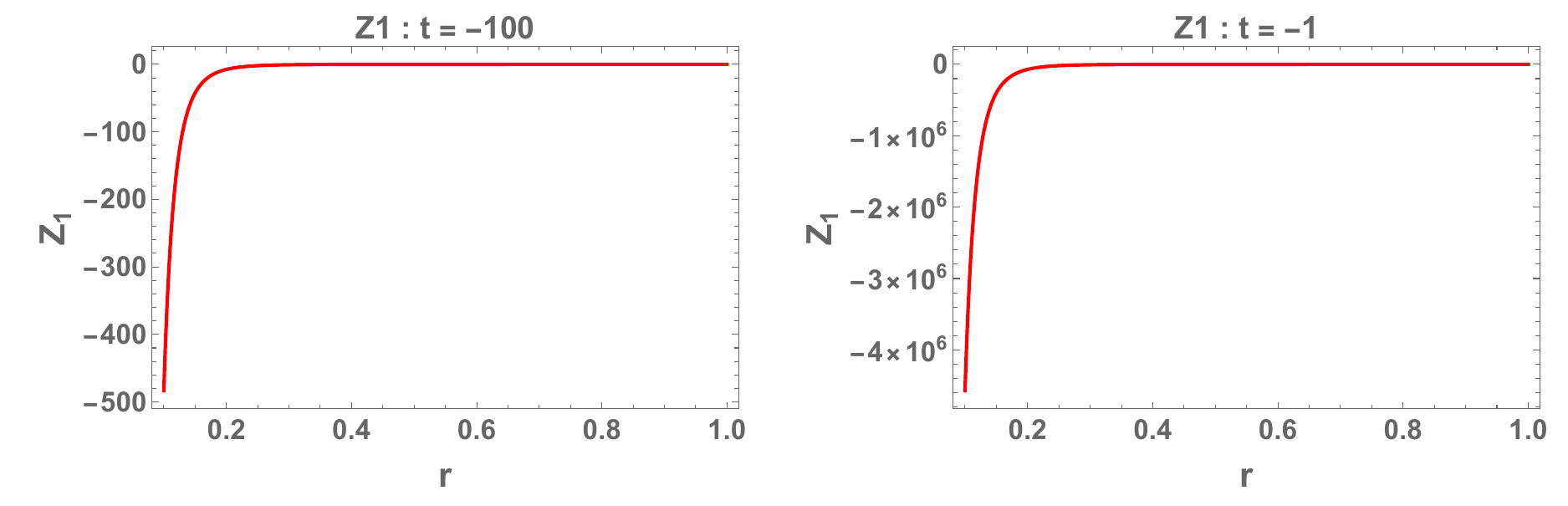}\label{fig:6a}}\\
\subfloat[]{\includegraphics[width=100mm]{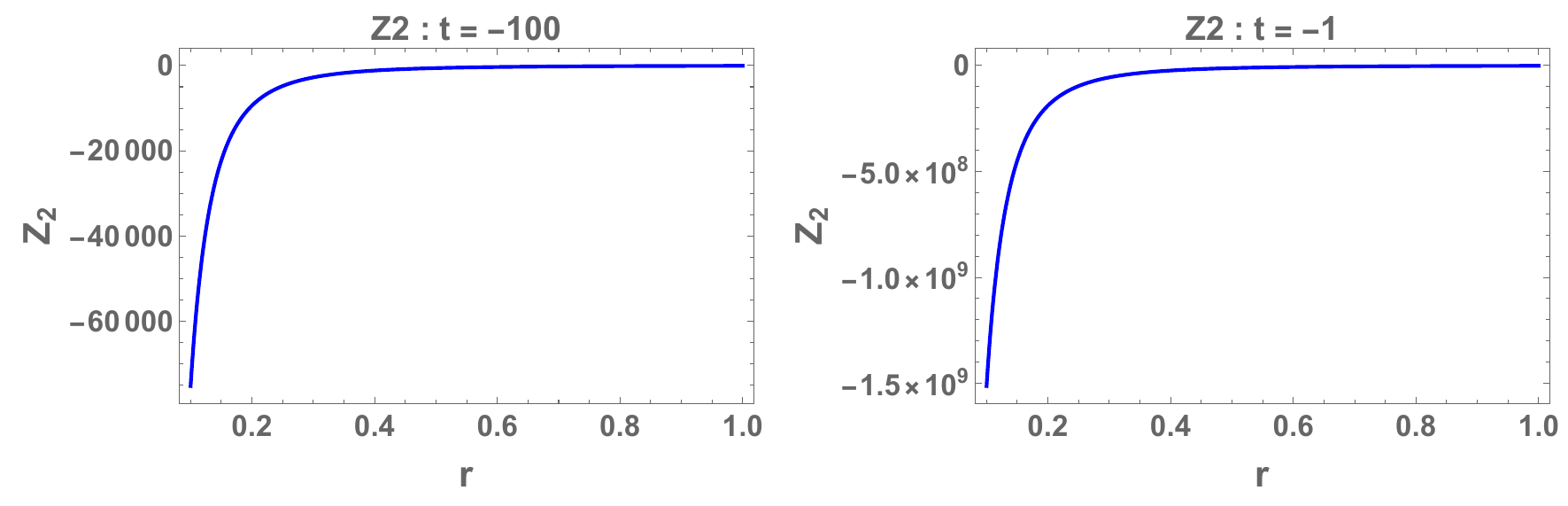}\label{fig:6b}}\\
\subfloat[]{\includegraphics[width=100mm]{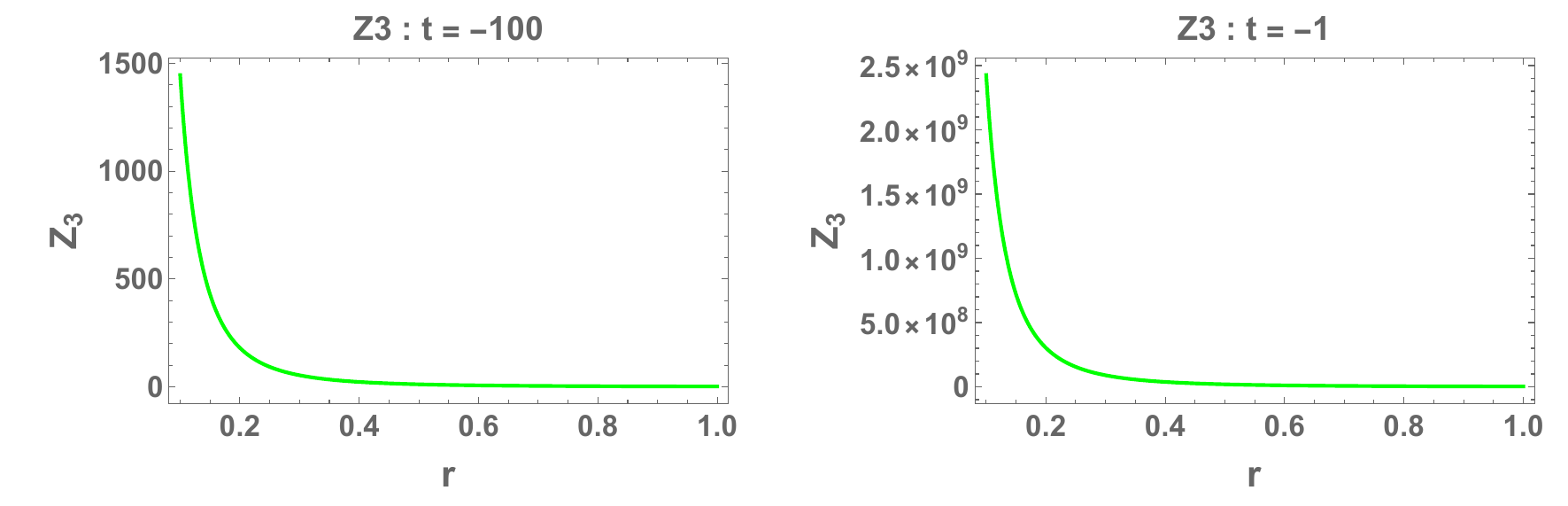}\label{fig:6c}}\\
\subfloat[]{\includegraphics[width=100mm]{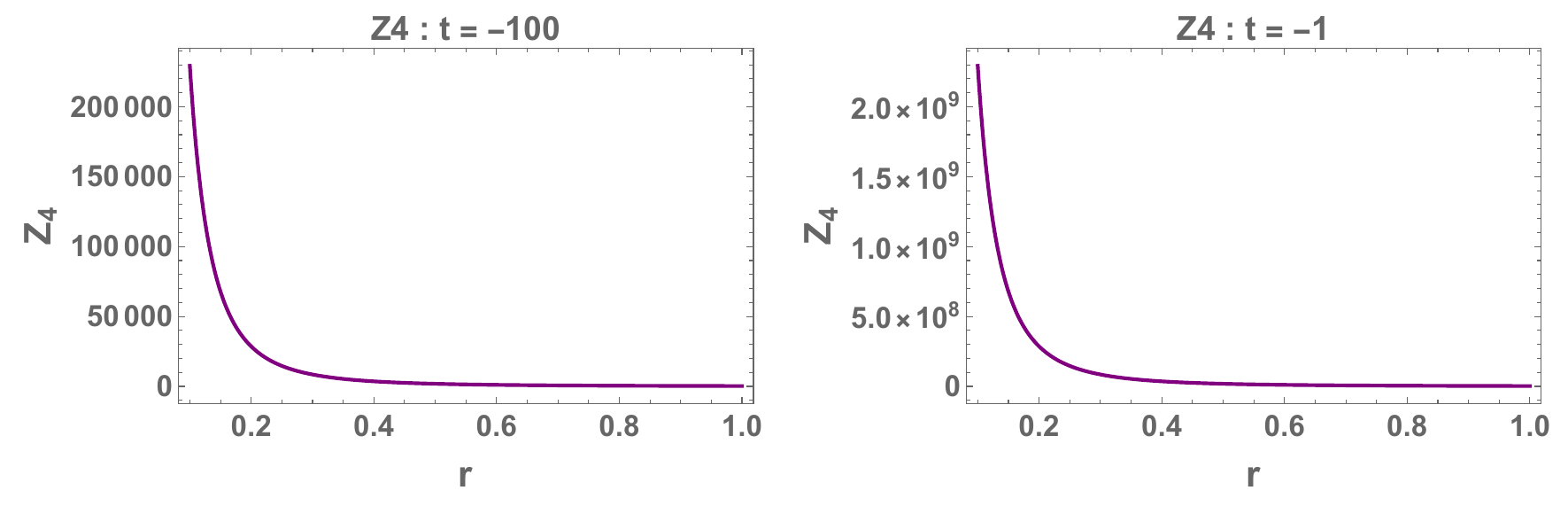}\label{fig:6d}}\\
\hspace{200mm}
\caption{Profiles of the individual contributions to the complexity factor at early time $t=-100$ (left) and late time $t=-1$ (right): (a) anisotropy $Z_{1}$, (b) density inhomogeneity $Z_{2}$, (c) dissipation $Z_{3}$, and (d) electric charge $Z_{4}$, for $H=\phi=a=1$.}
\label{fig:6}
\end{figure}

The $Y_{TF}$ is shown in Fig. (\ref{fig:5}) (right panel). It is positive in the entire stellar interior and reduces monotonically with increasing values of the radial coordinate. In order to comprehend the cause of this behavior, the different components contributing to this total quantity are analyzed separately through Fig. (\ref{fig:6}).

% \begin{figure}[!htbp]
% \centering
% \includegraphics[width=130mm]{z.pdf}
% \hspace{200mm}
% \caption{}
% \label{fig:6}
% \end{figure}
% \FloatBarrier

\FloatBarrier

\section{Summary and Conclusion}
\label{sec:6}
The present study lies in extending the shearing radiating collapse framework to the charged case and examining how the electromagnetic field modifies the boundary evolution, energy conditions, sound-speed behaviour, and complexity factor.
The contribution is therefore not the inclusion of charge alone, but its systematic incorporation into this particular exact solution and physical analysis.
In the current study, the charged gravitational collapse and anisotropic radiating star with non-zero shear and radial heat flux has been considered. The electromagnetic field was included using the Einstein–Maxwell field equations, where the charge was made to directly contribute to the matter variables, and boundary dynamics. With the choice of ($Y=rR(t)$) and assuming that ($R=at$), the charged boundary condition was transformed into a Riccati-type equation. After the transformation ($f=t/B$), the Riccati-type equation was solved for the special case of ($\Delta=0$).\\
From the physical analysis, the density and radial pressure are positive and decreasing towards the surface of the star; however, the tangential pressure is still negative, which ensures the anisotropic nature of the charged system. The heat flow is maximum in the central region of the star; however, the contribution due to electromagnetic field is maximum in the interior region of the star. The energy conditions ($E_{2}$) and ($E_{3}$) are satisfied in the entire space; however, ($E_{1}$) changes its sign, which implies that the requirement of the real eigenvalues is valid for some part of the space. The radial sound speed satisfies the causality condition at both early and late stages, while the tangential sound speed shows negative behaviour. However, the cracking function is negative in the whole stellar distribution, which shows potential stability against the cracking phenomenon. In addition, the complexity factor is positive and decreasing in nature; however, the electric charge provides an extra contribution to the complexity factor, along with anisotropy, density inhomogeneity, and heat dissipation. These findings indicate that electrical charge changes the internal dynamics and structure of the collapsing radiation.

\section*{Data availability statement}
\bigskip
My manuscript has no associated data.

%\section*{References}

\end{document}